\documentclass[pra,reprint]{revtex4-2}
\usepackage{graphicx,url}
\usepackage{amsmath,braket,bm}
\usepackage{multirow}
\usepackage{hyperref}

\renewcommand{\vec}[1]{{\bm{#1}}}

\begin{document}
\title{
  Quantum signatures of the mixed classical phase space\\
  for three interacting particles in a circular trap}
\author{D.J.\ Papoular}
\email[Electronic address: ]{david.papoular@cyu.fr}
\author{B.\ Zumer}
\affiliation{LPTM, UMR 8089 CNRS \& CY Cergy Paris Universit\'e,
  Cergy--Pontoise, France}
\date{\today}

\begin{abstract}
  We study theoretically two consequences of the mixed  classical phase
  space for three repulsively--interacting bosonic particles in a circular trap. First, we
  show that the energy levels of the corresponding quantum system are well described
  by a Berry--Robnik distribution. Second, we identify stationary quantum states whose density
  is enhanced along the stable classical periodic trajectories,
  and calculate their energies and wavefunctions using
  the semiclassical Einstein--Brillouin--Keller (EBK) theory.
  Our EBK results are in excellent agreement with our full--fledged finite--element numerics.
  We discuss the impact of discrete symmetries, including bosonic exchange symmetry, on these
  classically localized states. They
  are within experimental reach, and occur in the
  same range of energies as the quantum scar reported in our previous work
  [Phys.~Rev.~A \textbf{107}, 022217 (2023)].
\end{abstract}

\maketitle

\begin{figure*}
  \begin{minipage}{.32\textwidth}
    \includegraphics[width=\textwidth]
    {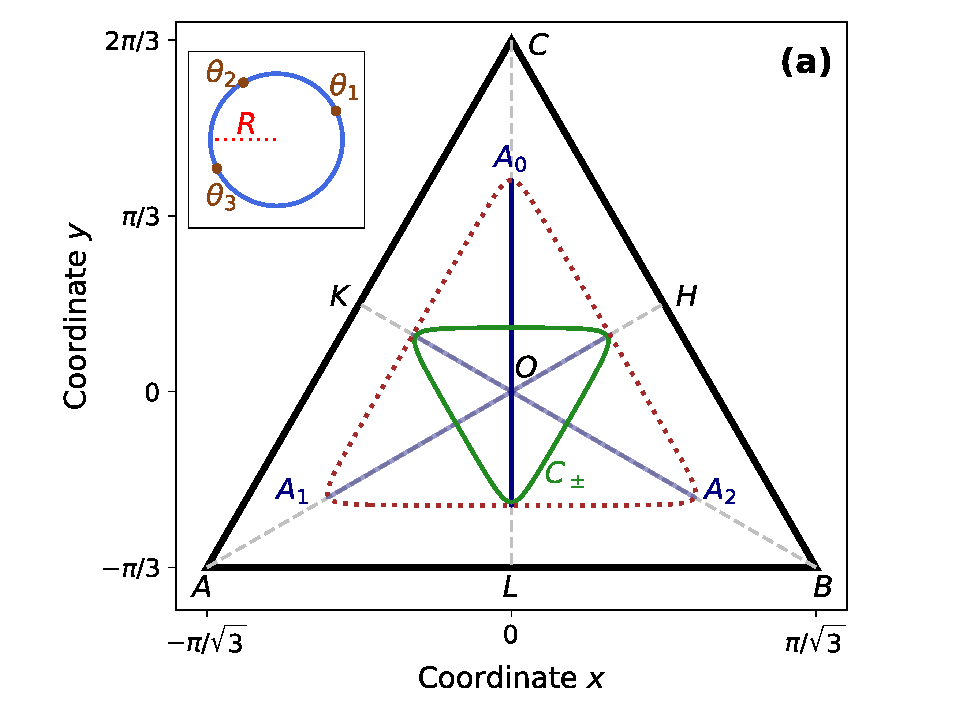}
  \end{minipage}
  \begin{minipage}{.32\textwidth}
    \includegraphics[width=\textwidth]
    {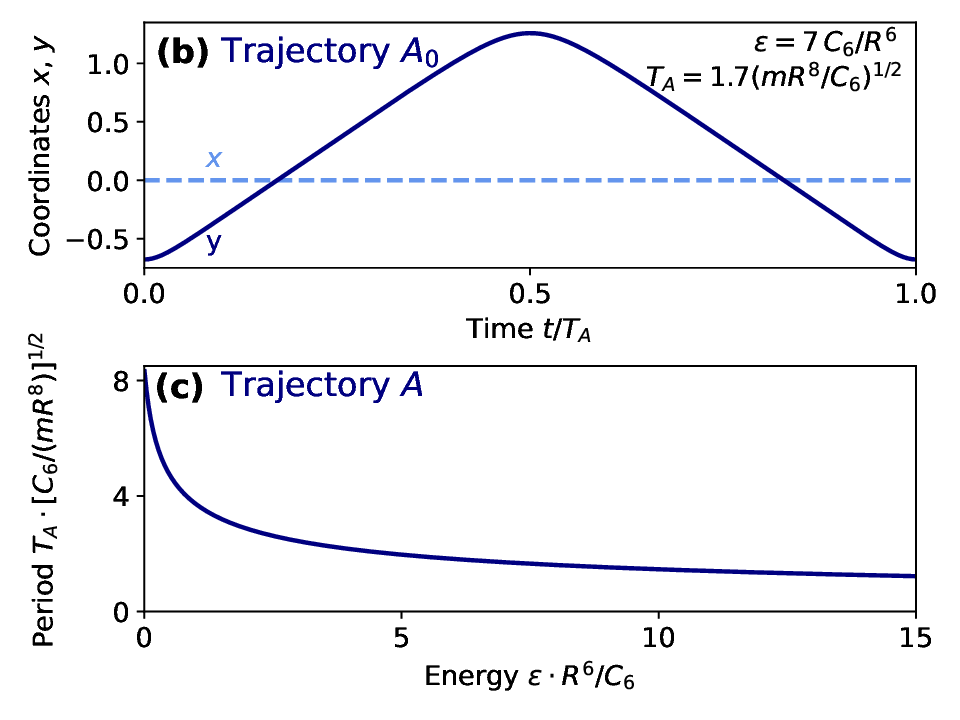}
  \end{minipage}
  \begin{minipage}{.32\textwidth}
    \includegraphics[width=\textwidth]
    {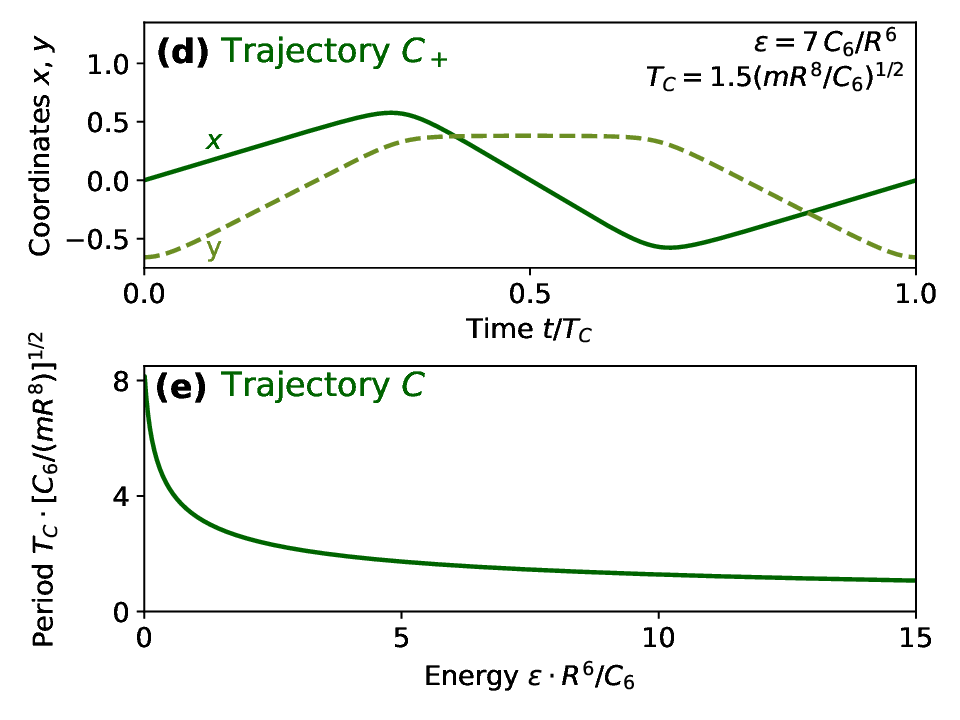}
  \end{minipage}
  \caption{
    \label{fig:pertrajAC}
    \textit{{(a)}}
    The periodic trajectories $A_0$, $A_1$, $A_2$ (straight blue lines)
    and $C_+$, $C_-$ (the closed green trajectory is followed anticlockwise for
    $C_+$ and clockwise for $C_-$),
    shown in the $(x,y)$ plane for the energy
    $\epsilon=7C_6/R^6$. The dotted brown line shows the classically accessible region.
    The inset shows the considered physical system: three interacting particles in a circular trap.
    \textit{{(b)}} Periodic trajectory $A_0$ as a function of time
    for $\epsilon=7C_6/R^6$ in terms of its coordinates
    $x(t)$ (solid line) and $y(t)$ (dashed line).
    \textit{{(c)}} Period $T_A(\epsilon)$ of trajectory $A$
    as a function of the energy $\epsilon$.
    Panels \textit{{(d)}} and \textit{{(e)}}
    show the corresponding quantities for trajectory $C_+$. Trajectories
    $A$ and $C$ are stable for the considered range of energies.}
\end{figure*}

\section{Introduction}
The suppression of ergodicity in quantum systems has long been under intense
scrutiny \cite[chap.~8]{haake:Springer2018}, and atomic systems are
very well suited to its investigation
Refs.~\cite[chap.~4]{stockmann:CUP1999}.
The mechanisms leading to it in many--body systems,
relying on e.g.\ integrability \cite{guan:RPP2022},
the presence of disorder \cite{abanin:RMP2019},
many--body scarring \cite{bernien:Nature2017,turner:NatPhys2018},
or periodic driving \cite{bluvstein:Science2021},
hold promises for
quantum information processing over long times, but  may
hinder cooling mechanisms \cite{brune:PRR2020}.

In the case of Hamiltonian systems,
comparing the quantum system to its classical analog has been very fruitful
in identifying such mechanisms \cite{bohigas:PhysRep1993}. Most classical systems
have a mixed phase space hosting both ergodic  and
non--ergodic trajectories. Ergodic trajectories densely cover a substantial fraction of
the energy surface; non--ergodic ones wind around tori found within the
Kolmogorov--Arnold--Moser (KAM) regions of phase space,
well described using KAM theory \cite[appendix 8]{arnold:Springer1989}.
Ergodicity in the quantum system may be suppressed in a phase space region corresponding
to classical ergodic motion, e.g.\ by a quantum scar \cite{heller:PRL1984}.
The quantum system is also known to exhibit regular levels reflecting the classical
non--ergodic trajectories \cite[Sec.~4]{bohigas:PhysRep1993}. These levels may be studied using the 
semiclassical Einstein--Brillouin--Keller (EBK) theory
\cite{keller:AnnPhys1958,percival:AdvChemPhys1977}. In contrast to the semiclassical
approaches applicable to the classically chaotic region, which mainly provide information
concerning the density of states \cite[chap.~17]{gutzwiller:Springer1990},
EBK theory applied to the classical KAM regions yields both quantum energy eigenvalues
and eigenfunctions
constructed from classically non--ergodic trajectories.
The full energy spectrum, including both the regular levels
to which EBK theory applies and the remaining levels related to
chaotic dynamics \cite[Sec.~5]{bohigas:PhysRep1993},  exhibits energy level statistics
which significantly deviate \cite{brody:LettNuovoCimento1973,berry:JPhysA1984}
from both the Poisson and Wigner
distributions  respectively
associated with classical integrability and chaos
\cite[chap.~16]{gutzwiller:Springer1990}.

Mixed classical phase spaces are relevant for the
description of many--body systems. 
The many--body scar
affecting the spin dynamics of a Rydberg atom chain
observed in Ref.~\cite{bernien:Nature2017} provides a recent example.
The classical analog system, whose construction is involved \cite{turner:PRX2021},
exhibits mixed phase space, and KAM regions play a key role in the many--body
quantum revivals \cite{michailidis:PRX2020}.
Motivated by these recent developments, we introduced in our previous
article \cite{papoular:PRA2023} the system of three interacting particles
in a circular trap. We analyzed this experimentally accessible system
through well--established theories applied to
a phase space whose dimension matches
the number of independent parameters introduced in Ref.~\cite{michailidis:PRX2020},
and identified a quantum scar affecting the motion of the atoms.

In this paper, we analyze the role of its mixed classical phase space.
First, we show that the parameters we investigated in Ref.~\cite{papoular:PRA2023}
fall within a range where the quantum energy level statistics are well
described by the Berry--Robnik distribution \cite{berry:JPhysA1984}.
Then, we identify quantum states whose probability density is enhanced
near stable classical periodic trajectories.
Using EBK theory, we characterize
their energy eigenvalues and explicitly construct their wavefunctions.
Our results are in excellent agreement with our full--fledged numerical solution
of the Schrödinger equation using the finite--element method.
We highlight the role of discrete symmetries, including bosonic echange symmetry,
and their observable consequence,
on the energies and wavefunctions of the considered localized
states.

We formulate our analysis in terms of trapped Rydberg atoms, made
accessible by  recent experimental advances \cite{cortinas:PRL2020,barredo:PRL2020}.
However, similar phenomena are expected to occur with systems
of magnetic atoms \cite{chomaz:RepProgPhys2023}
or polar molecules \cite{bohn:Science2017} exhibiting the same symmetries.
The classically--localized states  \cite[chap.~22]{heller:Princeton2018}
identified in the present paper occur for the same parameters and energy range
as the previously identified quantum scar \cite{papoular:PRA2023}.
One may address one effect or the other 
simply by changing the initial condition
defining the atomic motion. Hence, the simple, well--controlled
atomic system we are proposing offers an opportunity for
a detailed experimental comparison of the two effects.

The paper is organized as follows.
In Sec.~\ref{sec:system_summary}, we introduce the considered system,
and briefly summarize its properties described in detail in our previous
article \cite{papoular:PRA2023}.
In Sec.~\ref{sec:levelstats}, we show that its quantum energy levels
are well represented by the Berry--Robnik distribution.
In Sec.~\ref{sec:EBK_energies_wavefunctions}, we apply EBK theory to identify the
energy levels for the quantum states localized near stable periodic trajectories
and construct the corresponding EBK wavefunctions,
and we compare them to our finite--element numerical results.
In Sec.~\ref{sec:expprospects}, we discuss experimental prospects.
The article ends with the conclusive Sec.~\ref{sec:conclusion}.

\section{\label{sec:system_summary}
  The considered system}

The system we analyze has been introduced
in our previous article \cite{papoular:PRA2023}.
We briefly summarize its key features.

We consider three identical bosonic particles of mass $m$ in
a circular trap of radius $R$ (Fig.~\ref{fig:pertrajAC}(a),inset). We
assume that the interaction $v(d_{ij})$ between the particles $i$ and $j$
only depends on their distance $d_{ij}=2R|\sin[(\theta_i-\theta_j)/2]|$.
For circular Rydberg atoms whose electronic angular momenta are perpendicular
to the plane, $v(d_{ij})=C_6/d_{ij}^6$ with $C_6>0$. 
We introduce the Jacobi coordinates  
$x=[(\theta_1+\theta_2)/2-\theta_3+\pi]/\sqrt{3}$,
$y=(\theta_2-\theta_1)/2-\pi/3$,
$z=(\theta_1+\theta_2+\theta_3)/3-2\pi/3$, and
their conjugate momenta $p_x$, $p_y$, $p_z$ (which carry the unit of action).
Then, the Hamiltonian reads $H=p_z^2/(3mR^2)+H_{\mathrm{2D}}$, where 
\begin{equation}
  \label{eq:H2D}
  H_\mathrm{2D}=\frac{p_x^2+p_y^2}{4mR^2}+V(x,y)
  \ .
\end{equation}
Here, $V(x,y)=v(x,y)\,C_6/R^6$, with
\begin{multline}
  \label{eq:vxy}
  v(x,y)=[\sin^{-6}(\pi/3+y)+\sin^{-6}(\pi/3+x\sqrt{3}/2-y/2)\\
  +\sin^{-6}(\pi/3-x\sqrt{3}/2-y/2)
  ]/64-1/9,
\end{multline}
energies being measured from the minimum $V(\vec{0})$.
The Hamiltonian $H$ may be understood as describing either a classical system
or its quantum counterpart.
It is invariant under the point group
$C_{3v}$, generated by the rotation 
of order 3 about
the axis $(x=y=0)$ and the reflection in the plane $(x=0)$.
The free motion of the coordinate $z$ reflects the conservation of the total
angular momentum $p_z$. Once the latter is fixed,
the system is reduced to an effective point in the two--dimensional (2D)
plane $(x,y)$ within the equilateral triangle ABC of Fig.~\ref{fig:pertrajAC}(a),
in the presence of the potential $V(x,y)$.

From the quantum point of view, we seek the 3--atom
eigenstates of $H$ in the form
$\Psi_n(\theta_1,\theta_2,\theta_3)=\psi_n(\vec{r})\, e^{inz}$, where $\vec{r}=(x,y)$,
and $n$ is an integer setting the value of the quantized angular momentum $p_z$.
The wavefunction $\psi_n(\vec{r})$ is fully
determined by its values within the triangle $ABC$ and vanishes along $AB$, $BC$,
and $CA$.
The constraint
$\Psi_n(\theta_1,\theta_2,\theta_3)=\Psi_n(\theta_3-2\pi,\theta_1,\theta_2)$,
combining bosonic symmetry and angular periodicity, yields:
\begin{equation}
  \label{eq:psin_rotation}
  \psi_n(\mathcal{R}\vec{r})=\psi_n(\vec{r})\, e^{2in\pi/3}
  \ ,
\end{equation}
where $\mathcal{R}$
is the rotation of angle $2\pi/3$ about $O$ in the $(x,y)$ plane.
We sort the energy levels in terms of the three irreducible
representations $A_1$, $A_2$, $E$
of $C_{3v}$.
Owing to Eq.~(\ref{eq:psin_rotation}),
wavefunctions pertaining to
the one--dimensional (1D) representations $A_1$ or $A_2$ have $n=0$ modulo 3,
whereas those pertaining to the 2D representation $E$ have
$n\neq 0$ modulo $3$.

As in Ref.~\cite{papoular:PRA2023},
we set the ratio
$\eta=\hbar R^2/(mC_6)^{1/2}$ to $0.01$,
and we consider energies $\epsilon\sim 7C_6/R^6$.

\section{\label{sec:levelstats}
  Mixed classical phase space and quantum energy level statistics}

\subsection{Classical periodic trajectories}

We have characterized the periodic trajectories of the model of Eq.~(\ref{eq:H2D})
using our own C++ implementation of the numerical approach
of Ref.~\cite{baranger:AnnPhys1988}. We find three families of periodic trajectories
existing for all energies $\epsilon>0$: we label them $A$, $B$, $C$ in analogy with
those of the H\'enon--Heiles potential \cite{davies:Chaos1992}. We have
analyzed the unstable trajectories of family $B$ (i.e\ their Lyapunov exponent $>0$),
along with the quantum scar it yields,
in our previous article \cite{papoular:PRA2023}. By contrast,
the trajectories of families $A$ and $C$ are  stable for all considered energies
(i.e.\ their Lyapunov exponents $=0$).

For a given energy $\epsilon$, family A contains three straight--line trajectories
$A_0$, $A_1$, $A_2$, which follow the medians of the triangular configuration space,
and transform into one another under rotations of order 3. Family C
contains two trajectories $C_+$ and $C_-$, which are closed loops around
the center $O$: $C_+$ is followed anticlockwise and $C_-$ clockwise, and they transform
into each other under reflections about any of the three medians.
All five trajectories are represented in the $(x,y)$ plane on Fig.~\ref{fig:pertrajAC}(a).
The vertical trajectory $A_0$ and the trajectory $C_+$ are shown as functions of time
on Figs.~\ref{fig:pertrajAC}(b, d). Trajectories of a given
family have the same period as a function of energy $T_A(\epsilon)$ and $T_C(\epsilon)$:
these are plotted on Figs.~\ref{fig:pertrajAC}(c, e) and are both of the order of
$(mR^8/C_6)^{1/2}$ for $\epsilon\sim 7 C_6/R^6$.

\begin{figure}
  \includegraphics[width=\linewidth]
  {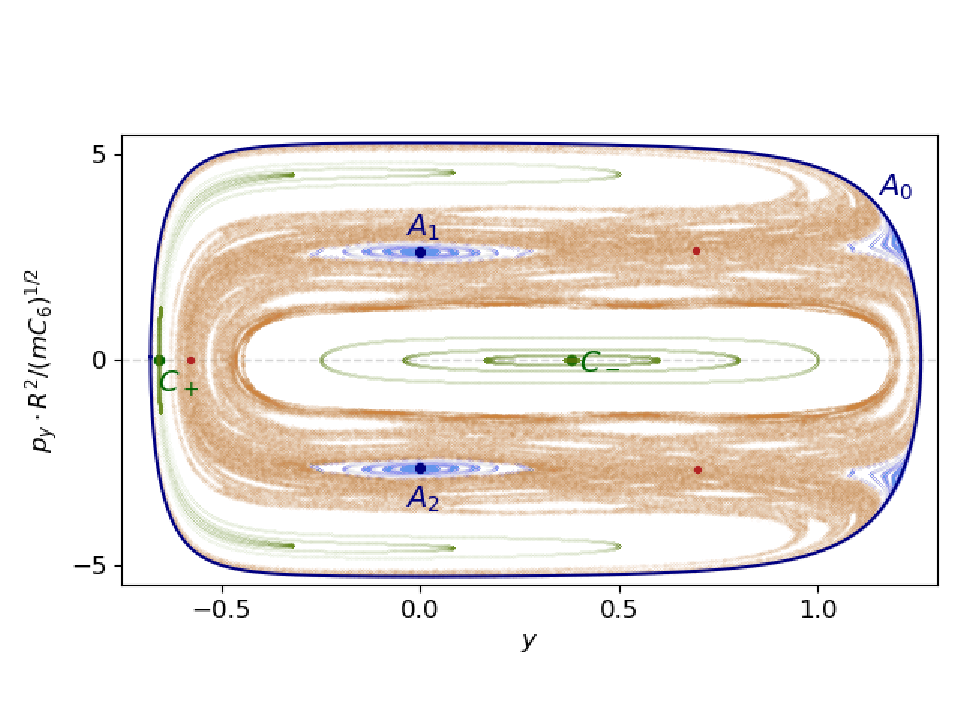}
  \caption{
    \label{fig:SoS_Rydberg}
    Surface of section for  Eq.~(\ref{eq:vxy}),
    with $p_z=0$,
    $\epsilon=7C_6/R^6$, $x=0$, and $p_x>0$. The periodic trajectory
    $A_0$ appears as the dark blue closed boundary of the figure.
    All other periodic trajectories appear as fixed points, shown in
    dark blue for $A_1$ and $A_2$;
    dark red for $B_1$, $B_2$, and $B_3$;
    and dark green for $C_+$ and $C_-$.
    The stable trajectories $A_i$ and $C_j$ are surrounded by
    (light blue and light green) tori; no tori are present
    near the unstable trajectories $B_k$.
    The $\approx 287000$ thin brown dots
    all belong to the same ergodic trajectory.
  }
\end{figure}

The simultaneous existence of stable  and unstable  periodic trajectories
signals that the classical system represented by $H_\mathrm{2D}$
is neither integrable nor fully chaotic:
its phase space 
is mixed. 
This is apparent on the surface of section of
Fig.~\ref{fig:SoS_Rydberg} \cite{papoular:PRA2023}.
There, the non--ergodic trajectories are represented
by the closed blue and green curves, which are sections in the two--dimensional
plane of the KAM tori \cite[appendix~8]{arnold:Springer1989}
surrounding the stable
trajectories $A$ and $C$.
We numerically find that the fraction of the surface of section not occupied
by tori is densely covered by the intersections from a single ergodic trajectory,
comprising the single ergodic zone visible on Fig.~\ref{fig:SoS_Rydberg}, within which
lie the 3 unstable trajectories of family $B$.

\subsection{
  Quantum energy level statistics}

\begin{figure}
  \includegraphics[width=\linewidth]
  {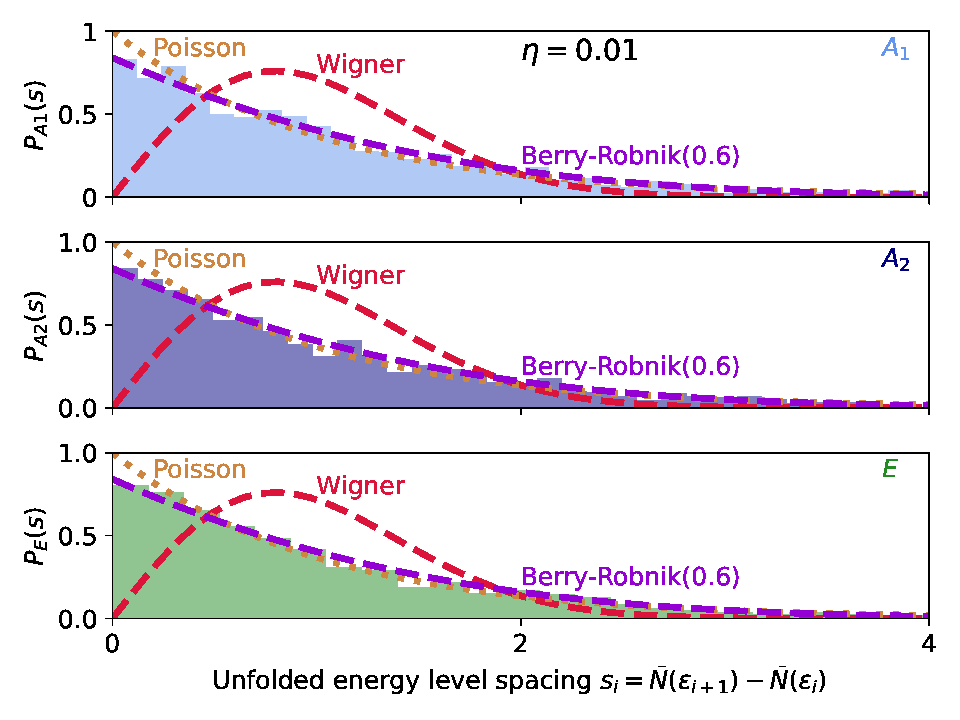}
  \caption{ \label{fig:levelspacings_Rydberg}
    The histograms show the distribution of unfolded energy level spacings
    $s_{r,i}=\bar{N}_r(\epsilon_i+1)-\bar{N}_r(\epsilon_i)$ for states
    belonging to the three irreducible
    representations $r=A_1$ (top), $A_2$ (center), $E$ (bottom),
    which are analyzed separately.
    They differ from the Poisson (dotted golden line) and Wigner (dashed red line).
    They are well represented by the Berry--Robnik distribution,
    assuming a single chaotic region in phase space, with 
    parameter $\rho_1=0.6$ for all three representations.
  }
\end{figure}
The quantum spectra of systems with mixed classical phase space satisfy
neither the Poisson nor the Wigner distribution
\cite[Sec.~16.8]{gutzwiller:Springer1990}.
We now verify this  for the model of Eq.~(\ref{eq:vxy}), and show that
its energy level statistics are well represented by
a Berry--Robnik distribution \cite{berry:JPhysA1984}.

We numerically solve the Schrödinger equation
for the Hamiltonian of  Eq.~(\ref{eq:H2D}) using the finite--element software
\textsc{FreeFEM} \cite{hecht:JNumerMath2012}. We calculate stationary states
belonging to
the three irreducible representations $A_1$, $A_2$, $E$
of the point group $C_{3v}$ separately.
We exploit 
discrete symmetries to reduce the configuration space to a triangle
which slightly exceeds 1/6 of the classically accessible region
for a given energy: details are given in our previous paper
\cite[Appendix 2]{papoular:PRA2023}.
We use a triangular mesh comprising $1000$ vertices along each edge.
We thus numerically obtain the energies and wavefunctions for slightly more than $1200$
consecutive energy levels for Representation $A_1$,
$1200$ levels for Representation $A_2$, and $1700$ non--degenerate levels for
Representation $E$, in energy windows centered on $7C_6/R^6$.

For each irreducible representation $r=A_1$, $A_2$, and $E$ of $C_{3v}$,
we introduce the integrated density
of states $N_r(\epsilon)$, which is the staircase--like function
giving the number of stationary quantum states 
whose energies are smaller than $\epsilon$
\cite[Sec.~16.2]{gutzwiller:Springer1990}.
We describe its smooth component  $\bar{N}_r(\epsilon)$ through its Weyl expansion,
accounting for discrete symmetries \cite{lauritzen:AnnPhys1995}.
We retain the leading--order term, proportional to $1/\hbar^2$, and
the first correction,
proportional to $1/\hbar$. We calculate the spacings
$s_{r,i}=\bar{N}_r(\epsilon_i+1)-\bar{N}_r(\epsilon_i)$ between consecutive
`unfolded' energies $\bar{N}_r(\epsilon_i)$
\cite[Sec.~5.4]{bohigas:PhysRep1993}.
We plot their distribution on Fig.~\ref{fig:levelspacings_Rydberg},
where it is seen to differ from  both the Poisson and the Wigner distributions
\cite[Secs.~16.3 \& 16.4]{gutzwiller:Springer1990},
as expected for a system with mixed classical phase space.

Figure \ref{fig:levelspacings_Rydberg} shows that
the distribution of unfolded energy level spacings
is well represented by the Berry--Robnik distribution \cite{berry:JPhysA1984},
assuming that a single chaotic region
in phase space contributes to the statistics, with the same parameter $\rho_1=0.6$
for all three representations.
Both the assumption of a 
single chaotic region and the value $\rho_1=0.6$,
representing
the fraction of the energy surface over which motion is regular, are compatible
with the surface of section of  Fig.~\ref{fig:SoS_Rydberg}.
The applicability of the Berry--Robnik distribution
hinges on the statistical independence of the regular and chaotic sequences of levels.
Counter--examples have been identified, e.g.\ the hydrogen atom in a magnetic field
\cite{wintgen:PRA1987}, and its numerical verification with billiards requires reaching
the deep semiclassical limit \cite{prosen:JPhysA1998}.
By contrast, our result provides a realization
of the Berry--Robnik distribution in an experimentally accessible system involving
smooth interatomic interactions rather than sharp billiard walls.

\section{\label{sec:EBK_energies_wavefunctions}
  Quantum stationary states localized near
  the classically stable periodic trajectories $A$ and $C$}

For the majority of the  stationary quantum states of the Hamiltonian $H_\mathrm{2D}$
that we have obtained numerically, the probability density $|\psi(x,y)|^2$
is not directly related to the periodic trajectories of types $A$ and $C$. Nevertheless,
we find multiple eigenstates whose probability density is enhanced along one or the
other of these
trajectories. Figures \ref{fig:TrajA_Schrod_EBK}(a,b) and
\ref{fig:TrajC_Schrod_EBK}(a,b) illustrate this phenomenon for trajectories
$A$ and $C$, respectively:
in each case, we show the probability density for
the quantum states closest to the energy $\epsilon=7C_6/R^6$.
This phenomenon superficially resembles
the quantum scars stemming from trajectory $B$ which we have identified in our previous
article \cite{papoular:PRA2023}. However, the quantum  states we consider in the present article
do not satisfy Heller's
definition for a quantum scar \cite[chap.~22]{heller:Princeton2018}. Indeed, in
stark contrast to the classically unstable trajectory $B$,
trajectories $A$ and $C$ are both classically stable. Hence, quantum mechanics
yields no qualitative change in the behavior of the system in their vicinity.
In this section, we illustrate this statement with two results. First, calculating
the energies of the quantum states related to trajectories $A$ and $C$ semiclassically,
we justify that they obey selection rules which we entirely explain in terms of the
symmetries of the classical KAM tori. Second, we construct semiclassical wavefunctions
for these quantum states. 
Our semiclassical results for both the energies and the wavefunctions are in excellent
agreement with our full quantum calculation.

\subsection{\label{sec:KAMtori_symmetries}
  Symmetries of the regular classical trajectories}

We first consider the regular classical trajectories in the KAM regions  of phase space
surrounding the stable periodic  trajectories of families $A$ and $C$.
Our numerical results show that the tori lying close to the periodic trajectories
inherit the discrete symmetry properties  of the corresponding
periodic  trajectories, namely:
\textit{(i)} A torus $T_A$ near
  the periodic trajectory of type $A$ invariant under the reflection $\mathcal{S}$
  exhibits reflection symmetry, i.e.\ 
  if the point $(\vec{\vec{r},\vec{p}})$ belongs to $T_A$, then so does
  $(\mathcal{S}\vec{r},\mathcal{S}\vec{p})$;
\textit{(ii)}  A torus $T_C$ near a periodic trajectory of type  $C$
  is  invariant under rotations $\mathcal{R}$ of order 3, i.e.\ 
  if the point $(\vec{\vec{r},\vec{p}})$ belongs to $T_C$, then so does
  $(\mathcal{R}\vec{r},\mathcal{R}\vec{p})$.

We justify properties \textit{(i)} and \textit{(ii)} through the following argument.
We rely on an approximation
introduced in Ref.~\cite[Sec.~4.1]{bohigas:PhysRep1993}:
we ignore narrow instability subregions
and approximate the whole KAM region by a set of concentric tori.
Our numerical results for the surface of section,
shown on Fig.~\ref{fig:SoS_Rydberg}, confirm that it is very well satisfied for
the inner tori, close to the periodic trajectories, which are of interest
in this work (it breaks down for the outer tori in the vicinity
of the ergodic zone, which we do not consider).
This allows for the introduction of local action--angle coordinates, valid within
this region. These are defined through the consistent
choice of fundamental frequencies $\vec{\omega}=(\omega_1,\omega_2)$
\cite[Sec.~III.E]{martens:JChemPhys1987} on each torus within the region.
Then, any conditionally--periodic trajectory $(\vec{r}(t),\vec{p}(t))$
winding around one such torus
may be written as a Fourier series \cite[\S 52]{landau1:BH1976}:
\begin{equation}
  \label{eq:condperiodic_Fourierseries}
  \vec{r}(t)=\sum_{\vec{k}} \vec{r}_{\vec{k}} \exp (i\vec{k}\cdot\vec{\omega}t), \quad
  \vec{p}(t)=2mR^2d\vec{r}/dt \ ,
\end{equation}
the sum being taken over all integer pairs $\vec{k}=(k_1,k_2)$.
The considered torus is uniquely determined by its actions $\vec{J}=(J_1,J_2)$,
which are given by \cite{percival:JPhysA1974}:
\begin{equation}
  \label{eq:Jalpha_Percival}
  J_\alpha=\sum_{\alpha'=1,2}\sum_{\vec{k}}
  k_\alpha |\vec{r}_{\vec{k}}|^2 k_{\alpha'} \omega_{\alpha'}
  \ .
\end{equation}

Let us justify statement \textit{(ii)}, concerning tori in the
vicinity of a periodic trajectory of type $C$.
We consider a point $(\vec{r},\vec{p})$ belonging to 
the KAM region surrounding trajectory $C_+$,
and the rotated point $(\vec{r}',\vec{p}')$ with
$\vec{r}'=\mathcal{R}_{2\pi/3}\vec{r}$ and $\vec{p}'=\mathcal{R}_{2\pi/3}\vec{p}$.
Trajectory $C_+$ is invariant
under rotations of order 3, so that $(\vec{r}',\vec{p}')$
also belongs to the same KAM region. We compare the two trajectories
$(\vec{q}(t),\vec{p}(t))$ and $(\vec{q}'(t),\vec{p}'(t))$ obtained
from the initial conditions $(\vec{r},\vec{p})$ and $(\vec{r}',\vec{p}')$.
Their Fourier components $\vec{r}_{\vec{k}}$ and $\vec{r}'_{\vec{k}}$,
defined by Eq.~(\ref{eq:condperiodic_Fourierseries}), satisfy
$\vec{r}'_{\vec{k}}=\mathcal{R}_{2\pi/3}\vec{r}_{\vec{k}}$, so that
$|\vec{r}'_{\vec{k}}|=|\vec{r}_{\vec{k}}|$. According to Eq.~(\ref{eq:Jalpha_Percival}),
the actions $J_\alpha$ only depend on the modulus $|\vec{r}_{\vec{k}}|$, hence,
they are the same for both trajectories. Therefore, the points
$(\vec{r},\vec{p})$ and $(\vec{r}',\vec{p}')$ belong to the same torus $T_{C_+}$.
Statement \textit{(i)} may be justified similarly.

\subsection{
  \label{sec:EBK_levels}
  EBK quantization: energy levels}

In this section, we obtain semiclassical predictions for the energies of the quantum levels 
related to trajectories $A$ and $C$, which are in excellent agreement with the values
obtained through our numerical solution of the Schrödinger equation
(see Figs.~\ref{fig:EBK_levels_TrajA}(b) and \ref{fig:EBK_levels_TrajC}(c)).
We also explain quasidegeneracies and derive selection rules, both of which are direct
consequences of the discrete symmetries
of the KAM tori presented in Sec.~\ref{sec:KAMtori_symmetries} above.

Our semiclassical description relies on
Einstein--Brillouin--Keller (EBK) theory \cite{keller:AnnPhys1958},
accounting for the Maslov phase corrections \cite[\S 7]{maslov:Reidel1981}.
This theory generalizes the Wentzel--Kramers--Brillouin approach \cite[\S 48]{landau3:BH1977}
to the quantization of regular classical motion
with more than one degree of freedom \cite{percival:AdvChemPhys1977}.
We use our own implementation as a Python script of the EBK approach, based on
Refs.~\cite{martens:JChemPhys1985,martens:JChemPhys1987}, which hinges on
the representation of conditionally--periodic motion in terms of the Fourier
series of Eq.~(\ref{eq:condperiodic_Fourierseries}). We integrate classical trajectories
over time intervals of lengths up to $t_\mathrm{max}=3700(mR^8/C_6)^{1/2}$ and keep
up to $3200$ terms in Eq.~(\ref{eq:condperiodic_Fourierseries}).

We now characterize
the quantum stationary states localized near
the classically stable trajectories $A$ and $C$.
In Sections \ref{sec:energylevels_TrajA} and \ref{sec:energylevels_TrajC} below,
we derive the EBK energies for these states, considered as eigenstates of $H_{\mathrm{2D}}$,
whose wavefunctions depend on $\vec{r}=(x,y)$.
In Section \ref{sec:role_angularmomentum}, we analyze the role of angular momentum
so as to discuss the stationary states of the three--particle Hamiltonian $H$,
whose wavefunctions depend on $(x,y,z)$.

\subsubsection{ \label{sec:energylevels_TrajA}
  Quantum states localized near trajectory A}

\begin{figure*}
  \includegraphics[width=.32\linewidth]
  {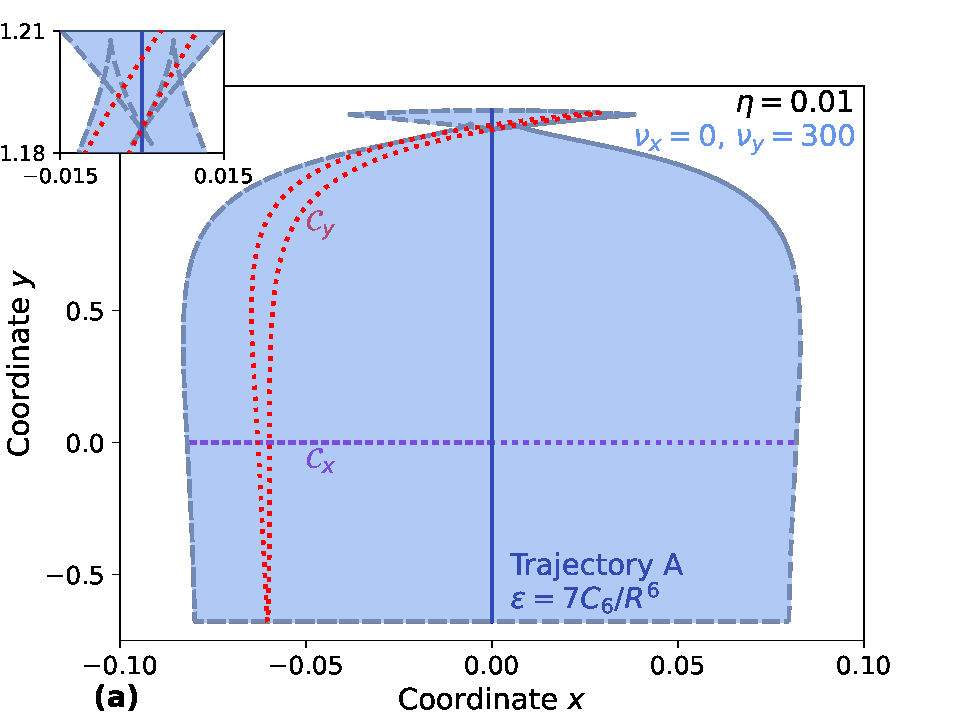}
  \includegraphics[width=.32\linewidth]
  {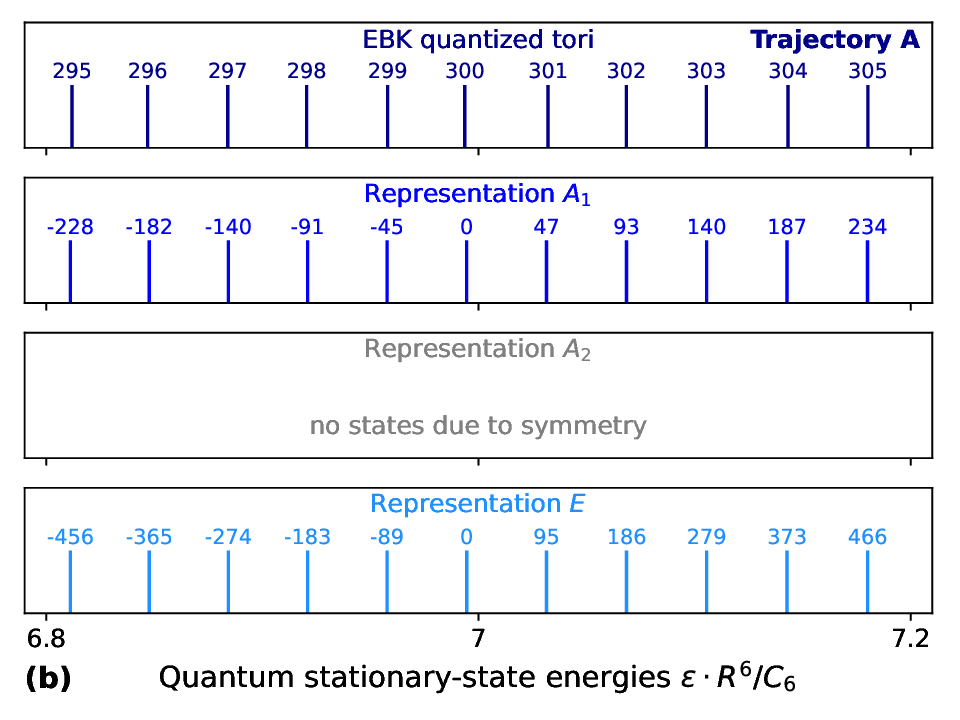}
  \includegraphics[width=.32\linewidth]
  {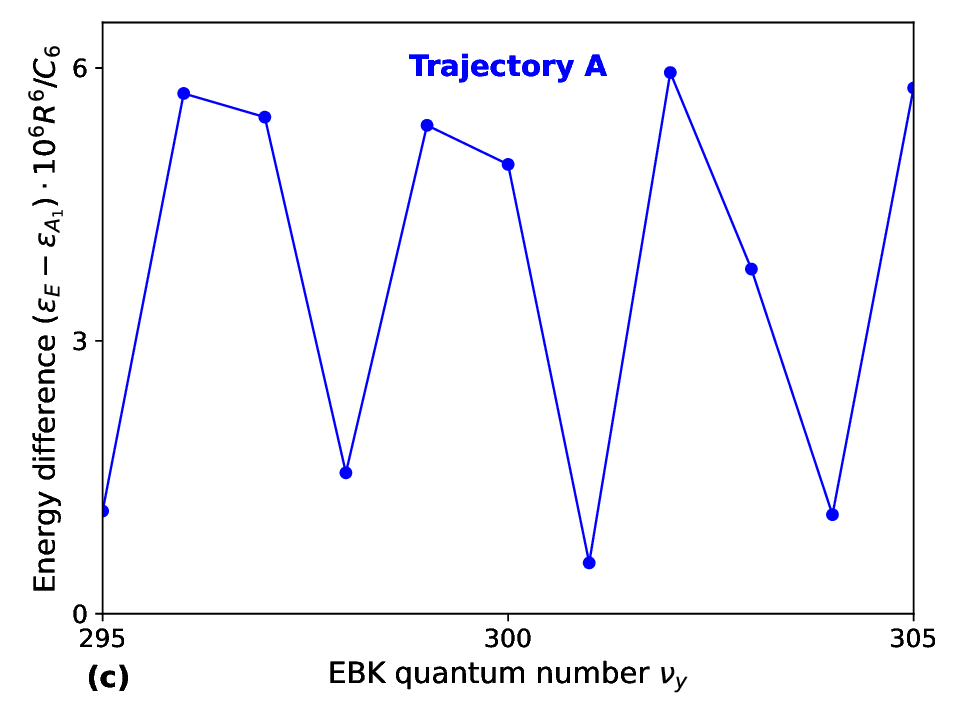}
  \caption{
    \label{fig:EBK_levels_TrajA}
    \textit{{(a)}} Classical trajectory $A$
    (solid dark blue)
    for the energy $\epsilon=7C_6/R^6$,
    the nearest--energy trajectory satisfying Eq.~(\ref{eq:EBKconds_TrajA})
    for $\eta=0.01$ (densely covering the light blue area),
    and two independent circuits
    $\mathcal{C}_x$ (dotted purple) and $\mathcal{C}_y$ (dotted red)
    circling the torus, in terms of which the quantum numbers are
    $\nu_x=0$ and $\nu_y=300$. The dashed gray lines show the caustics
    of this trajectory.
    The top left inset zooms in on the narrow region near $(x=0,y=1.2)$
    to reveal the self--intersection of the caustics.
    \textit{{(b)}} Top panel: energies of the EBK wavefunctions
    for $\nu_x=0$ and $295\leq \nu_y \leq 305$. Center and bottom panels:
    energies of the corresponding quasidegenerate
    quantum stationary states belonging
    to representations $A_1$ (center) and $E$ (bottom), obtained through our
    finite--element numerical calculations.
    Because of the torus symmetries, there are no states in representation
    $A_2$ corresponding to the EBK quantum numbers $(\nu_x=0,\nu_y)$.
    The integers in the center and
    bottom panels specify the relative state indices within each representation,
    $\Delta\nu^{A_1}$ and $\Delta\nu^{E}/2$,
    with respect to the quantum state related to Trajectory $A$ whose
    energy is closest to $7C_6/R^6$.
    \textit{{(c)}}
    Small energy differences between the quasidegenerate states of
    representations $A_1$ and $E$. 
  }
\end{figure*}

For a given energy $\epsilon$,
the three periodic trajectories $A_0$, $A_1$, and $A_2$ (see Fig.~\ref{fig:pertrajAC}(a))
and the tori surrounding them are mapped one onto the other through the rotations 
$\mathcal{R}$ and $\mathcal{R}^{-1}$. Hence, we focus on the vertical trajectory $A_0$.
In Eq.~(\ref{eq:condperiodic_Fourierseries}),
we choose the fundamental frequencies $\vec\omega=(\omega_1,\omega_2)$
as in Ref.~\cite[Fig.~8(b)]{martens:JChemPhys1987}. This leads to the independent
circuits $\mathcal{C}_x$ and $\mathcal{C}_y$ on Fig.~\ref{fig:EBK_levels_TrajA}(a).
Calculating their Maslov indices \cite[Sec.~II.C]{percival:AdvChemPhys1977},
we obtain the EBK quantization condition for
the tori near trajectory $A$:
\begin{equation}
  \label{eq:EBKconds_TrajA}
  I_x=\hbar(\nu_x+1/2)
  \text{ and }
  I_y=\hbar(\nu_y+1)
  \ ,
\end{equation}
where $I_{x,y}$ are the action integrals for the circuits $\mathcal{C}_{x,y}$, 
$\hbar$ is the reduced Planck's constant, and the integers $\nu_{x,y}\geq 0$ are the
EBK quantum numbers. The action $I_x\geq\hbar/2$, so that the periodic trajectory $A_0$
itself does not satisfy Eq.~(\ref{eq:EBKconds_TrajA}).
The  tori satisfying Eq.~(\ref{eq:EBKconds_TrajA})
which are closest to trajectory $A_0$ 
are those with $\nu_x=0$:
the corresponding energies within a window centered on $\epsilon=7C_6/R^6$
are shown on the top line of Fig.~\ref{fig:EBK_levels_TrajA}(b).
We compare them to the energies of the stationary quantum states
of $H_\mathrm{2D}$ belonging to representations $A_1$ and $E$
localized near the trajectories $A_0$, $A_1$, and $A_2$,
obtained through our finite--element calculations (see Fig.~\ref{fig:TrajA_Schrod_EBK}(a,b)).
These are shown on Fig.~\ref{fig:EBK_levels_TrajA}(b), middle and bottom lines,
and are in excellent agreement with the EBK results.

Figure~\ref{fig:EBK_levels_TrajA}(b) reveals that each EBK energy
corresponds to 
quasidegenerate quantum states pertaining to representations $A_1$  and $E$.
Furthermore, no quantum stationary states pertaining to
representation $A_2$ exhibit density profiles similar to
Fig.~\ref{fig:TrajA_Schrod_EBK}(a,b). Both of these properties
follow from the symmetries of the regular trajectories
identified in Sec.~\ref{sec:KAMtori_symmetries} above, through
a mechanism identified in Refs.~\cite{leopold:JPhysA1982} and
\cite[Sec.~4.2]{bohigas:PhysRep1993}
in the case where the discrete symmetry
at play had order 2. The system we consider provides examples of
the same phenomenon involving $C_{3v}$ symmetry, as we now show.

We consider the EBK wavefunction  $\psi_\mathrm{EBK}(\vec{r})$,
corresponding to a torus in the vicinity of trajectory $A_0$,
with the energy $\epsilon_\mathrm{EBK}$, satisfying
Eq.~(\ref{eq:EBKconds_TrajA}) with $\nu_x=0$. This torus is invariant
under the reflection $\mathcal{S}$ about the vertical axis $x=0$.
Therefore, as shown in \cite[Sec.~4.2]{bohigas:PhysRep1993}:
\begin{equation}
  \label{psiEBK_TrajA_symmetry}
  \psi_\mathrm{EBK}(\mathcal{S}\vec{r})
  =(-1)^{\nu_x}\psi_\mathrm{EBK}(\vec{r})
  =\psi_\mathrm{EBK}(\vec{r})
  \ .
\end{equation}
The EBK wavefunction $\psi_\mathrm{EBK}$ reflects the symmetry of the corresponding
classical torus, but does not automatically satisfy
the symmetry requirements of any representation.
We now project it 
onto the irreducible
representations \cite[\S 94]{landau3:BH1977}
$A_1$, $A_2$, and $E$.
This yields three linearly independent wavefunctions,
$\psi_{\mathrm{EBK}}^{A_1}$ and $\psi_{\mathrm{EBK}}^{E,\pm}$,
pertaining to the representations $A_1$
and $E$, corresponding to the same semiclassical energy.
In terms of kets $\ket{\psi}$,
with $\braket{\vec{r}|\mathcal{R}|\psi}=\psi(\mathcal{R}^{-1}\vec{r})$
and $\braket{\vec{r}|\mathcal{S}|\psi}=\psi(\mathcal{S}\vec{r})$,
they read:
\begin{equation}
  \label{eq:psiEBK_TrajA_A1_E}
  \begin{cases}
    \ket{\psi_{\mathrm{EBK}}^{A_1}}
    &=
      \alpha_{A_1}
      (1+\mathcal{R}+\mathcal{R}^{-1})
      \ket{\psi_{\mathrm{EBK}}}
    \\
    \ket{\psi_{\mathrm{EBK}}^{E,+}}
    &=
      \alpha_{E}
      (1+j^*\: \mathcal{R}+j\:\mathcal{R}^{-1})
      \ket{\psi_{\mathrm{EBK}}}
      \ ,
    \\
    \ket{\psi_{\mathrm{EBK}}^{E,-}}
    &=
      \alpha_{E}
      (1+j\:\mathcal{R}+j^*\: \mathcal{R}^{-1})
      \ket{\psi_{\mathrm{EBK}}}
      \ ,    
  \end{cases}
\end{equation}
In Eq.~(\ref{eq:psiEBK_TrajA_A1_E}), $\alpha_{A_1,E}$ 
are normalization coefficients,
and $j=e^{2i\pi/3}$.
We have used the relations $\mathcal{SRS}=\mathcal{R}^{-1}$
and Eq.~(\ref{psiEBK_TrajA_symmetry}).
The states $\ket{\psi_{\mathrm{EBK}}^{A_1}}$ and
$\ket{\psi_{\mathrm{EBK}}^{E,\pm}}$ satisfy
$\mathcal{R}\ket{\psi_{\mathrm{EBK}}^{A_1}}=\ket{\psi_{\mathrm{EBK}}^{A_1}}$,
$\mathcal{R}\ket{\psi_{\mathrm{EBK}}^{E,\pm}}=\pm j\ket{\psi_{\mathrm{EBK}}^{E,\pm}}$
and
$\ket{\psi_{\mathrm{EBK}}^{E,-}} =\mathcal{S}\ket{\psi_{\mathrm{EBK}}^{E,+}}$.
The component of $\psi_\mathrm{EBK}$ pertaining to representation $A_2$,
proportional to
$(1+\mathcal{R}+\mathcal{R}^{-1})(1-\mathcal{S})\ket{\psi_{\mathrm{EBK}}}$,
is $0$
because of Eq.~(\ref{psiEBK_TrajA_symmetry}).

\subsubsection{ \label{sec:energylevels_TrajC}
  Quantum states localized near trajectory C}

\begin{figure*}
  \includegraphics[width=.32\linewidth]
  {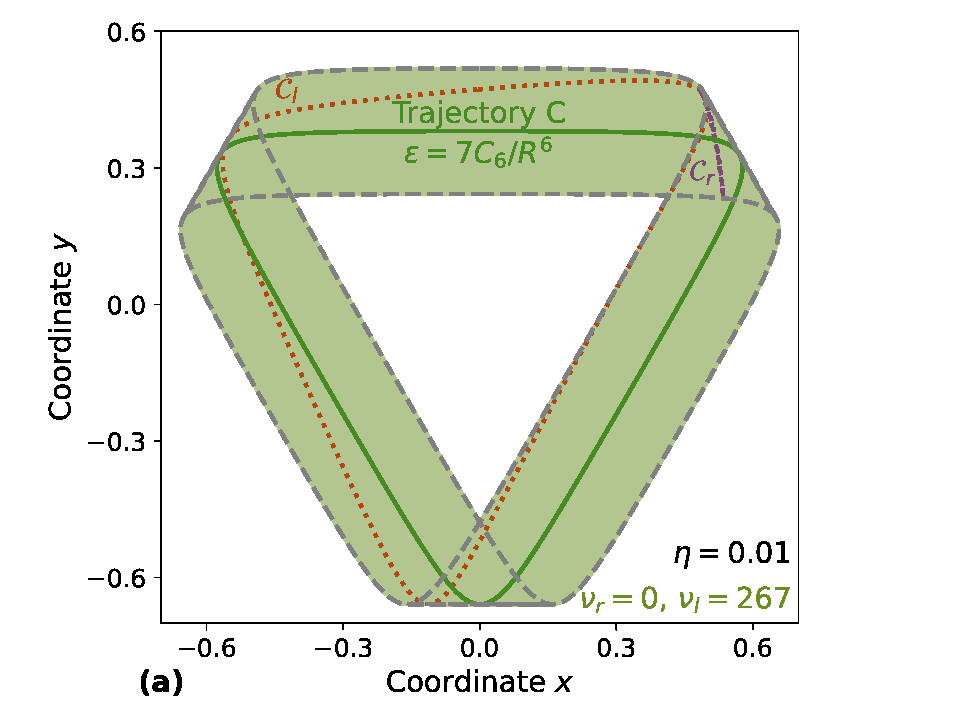}
  \includegraphics[width=.32\linewidth]
  {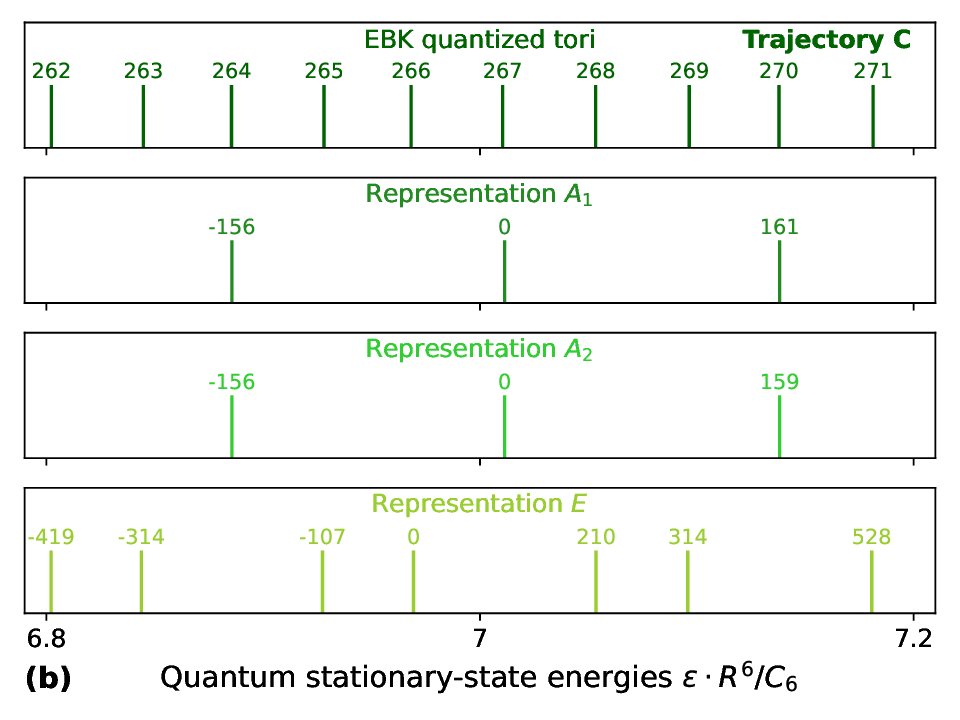}
  \includegraphics[width=.32\linewidth]
  {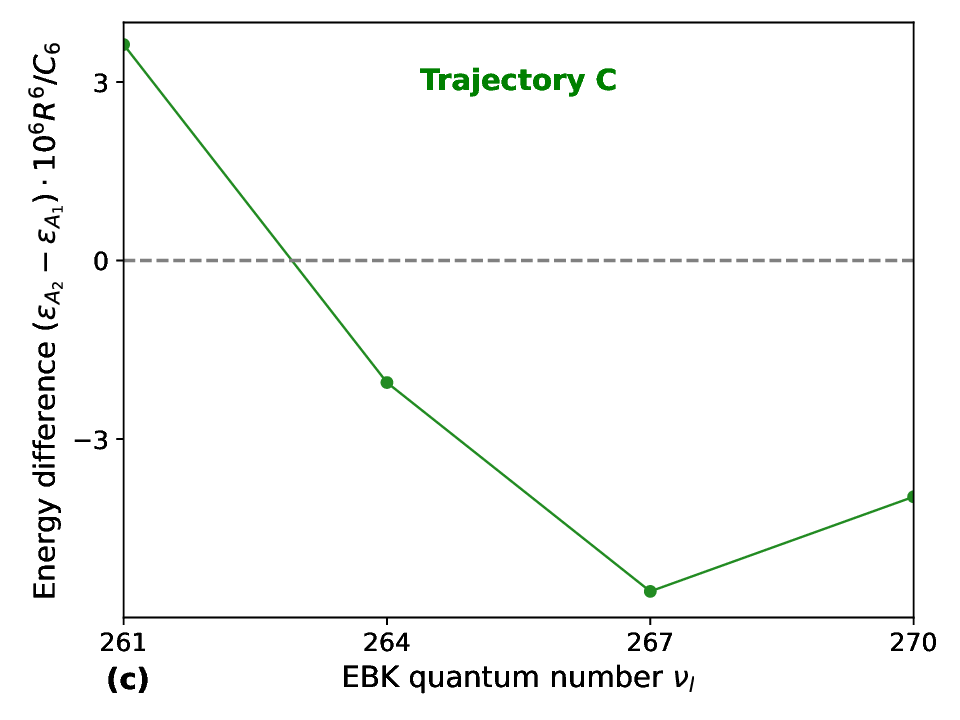}
  \caption{
    \label{fig:EBK_levels_TrajC}
    \textit{{(a)}} Classical trajectory $C$ (solid dark green)
    for the energy $\epsilon=7C_6/R^6$,
    the nearest--energy trajectory satisfying Eq.~(\ref{eq:EBKconds_TrajC})
    for $\eta=0.01$ (densely covering the light green area),
    and two independent circuits $\mathcal{C}_r$ (purple) and $\mathcal{C}_l$ (red)
    circling the torus, in terms of which the quantum numbers are
    $\nu_r=0$, $\nu_l=267$.
    The dashed gray lines show the caustics
    of this trajectory, which self--intersect in the top left, top right,
    and bottom regions.
    \textit{{(b)}} Top panel: energies of the EBK wavefunctions
    for $\nu_r=0$ and $262\leq \nu_l \leq 271$. Three lower panels:
    energies of the corresponding
    quantum eigenstates belonging
    to representations $A_1$, $A_2$, and $E$, obtained through our
    finite--element numerical calculations.
    States in representations $A_1$ and $A_2$ exhibit quasidegeneracies
    and correspond to the EBK quantum numbers $\nu_r=0$, $\nu_l= 0\text{ modulo 3}$;
    each EBK torus with quantum numbers
    $\nu_r=0$, $\nu_l\neq 0$ modulo 3
    yields two degenerate states in representation $E$.
    The integers specify the relative state indices within each representation,
    $\Delta\nu^{A_1}$, $\Delta\nu^{A_2}$ and $\Delta\nu^{E}/2$,
    with respect to the quantum state related to trajectory $C$ whose
    energy is closest to $7C_6/R^6$.
    \textit{{(c)}}
    Small energy differences between the quasidegenerate
    states of representations $A_1$ and $A_2$.
  }
\end{figure*}

We proceed as in Sec.~\ref{sec:energylevels_TrajA}.
For a given energy $\epsilon$,
the two periodic trajectories $C_+$ and $C_-$ (see Fig.~\ref{fig:pertrajAC}(a))
and the tori surrounding them are mapped onto each other through the reflection $\mathcal{S}$.
Hence, we focus on the trajectory $C_+$. In Eq.~(\ref{eq:condperiodic_Fourierseries}),
we choose the fundamental frequencies $\vec{\omega}=(\omega_1,\omega_2)$ as in
Ref.~\cite[Fig.~8(a)]{martens:JChemPhys1987}, leading to the independent circuits
$\mathcal{C}_r$ and $\mathcal{C}_l$ on Fig.~\ref{fig:EBK_levels_TrajC}(a). Calculating
their Maslov indices, we obtain the EBK quantization
condition for the tori near trajectory $C$:
\begin{equation}
  \label{eq:EBKconds_TrajC}
  I_r=\hbar(\nu_r+1/2)
  \text{ and }
  I_l=\hbar(\nu_l+1/2)
  \ ,
\end{equation}
where $I_{r,l}$ are the action integrals for the circuits $\mathcal{C}_{r,l}$,
and the integers $\nu_{r,l}\geq 0$ are the EBK quantum numbers.
The trajectory $C_+$
does not satisfy Eq.~(\ref{eq:EBKconds_TrajC}).
The tori satisfying it which are closest to $C_+$ 
are those with $\nu_r=0$: their energies
are shown on the top line of Fig.~\ref{fig:EBK_levels_TrajC}(b).
We compare them to the energies of the stationary quantum states
of $H_\mathrm{2D}$ belonging to representations $A_1$, $A_2$, and $E$
localized near the trajectories $C_+$ and $C_-$,
obtained through our finite--element calculations (see Fig.~\ref{fig:TrajC_Schrod_EBK}(a,b)).
These are shown on the three lower lines of Fig.~\ref{fig:EBK_levels_TrajC}(b),
and are in excellent agreement with the EBK results.

Figure \ref{fig:EBK_levels_TrajC}(b) shows that each EBK energy with
$\nu_r=0$ and $\nu_l=0$ modulo 3 corresponds to two quasidegenerate quantum states 
pertaining to representations $A_1$ and $A_2$. By contrast,
each EBK energy with
$\nu_r=0$ and $\nu_l\neq 0$ modulo 3 corresponds to
two exactly degenerate quantum states 
spanning a representation $E$.
As for the states localized near trajectory $A$
(see Sec.~\ref{sec:energylevels_TrajA} above), these properties follow from
the symmetries of the regular trajectories (Sec.~\ref{sec:KAMtori_symmetries}).
These are different from the symmetries of the tori surrounding trajectory $A$,
leading to different selection rules,
which we now derive.

We consider the EBK wavefunction $\chi_{\mathrm{EBK}}(\vec{r})$, corresponding to
a torus in the vicinity of trajectory $C_+$, with the energy $\epsilon_{\mathrm{EBK}}$,
satisfying Eq.~(\ref{eq:EBKconds_TrajC}) with $\nu_r=0$. This torus is invariant
under the rotation $\mathcal{R}$. A straightforward generalization of the
argument in Ref.~\cite[Sec.~4.2]{bohigas:PhysRep1993} to symmetry operations of order $3$
leads to
$\chi_{\mathrm{EBK}}(\mathcal{R}\vec{r})=j^{\nu_l}\: \chi_{\mathrm{EBK}}(\vec{r})$.
We now project $ \chi_{\mathrm{EBK}}$ onto the irreducible representations
$A_1$, $A_2$, and $E$. For each $\nu_l$, this yields two linearly independent, degenerate
EBK wavefunctions.
If $\nu_l=0$ modulo 3, the non--vanishing
wavefunctions pertain to representations $A_1$ and $A_2$:
\begin{equation}
  \label{eq:chiEBK_TrajC_A1_A2}
  \ket{\chi_{\mathrm{EBK}}^{A_1,A_2}}
  =
  \beta_{A_1,A_2}(1\pm \mathcal{S})\ket{\chi_{\mathrm{EBK}}}
  \ ,
\end{equation}
with $\beta_{A_1,A_2}$ being two normalization factors, whereas the component
along $E$ vanishes. By contrast, if $\nu_l\neq 0$ modulo 3,
the components along $A_1$ and $A_2$ vanish, whereas the
two non--vanishing wavefunctions
$\ket{\chi_{\mathrm{EBK}}^{E,\pm}}$ span a representation $E$. For $\nu_l=-1$ modulo 3,
$\ket{\chi_{\mathrm{EBK}}^{E,+}}=\ket{\chi_{\mathrm{EBK}}}$
and
$\ket{\chi_{\mathrm{EBK}}^{E,-}}=\mathcal{S}\ket{\chi_{\mathrm{EBK}}}$,
and the opposite assignment  holds for $\nu_l=+1$ modulo 3.

\subsubsection{ \label{sec:role_angularmomentum}
  The role of angular momentum}
To discuss
the three--particle eigenstates of $H$ in terms of the  eigenstates of $H_\mathrm{2D}$
identified in Secs.~\ref{sec:energylevels_TrajA} and \ref{sec:energylevels_TrajC},
we now analyze the role of angular momentum.

We first consider quantum states localized near the periodic trajectories of family $A$.
The two states $\psi^{E,\pm}_{\nu_y}(\vec{r})$ obtained for a given $\nu_y$,
are exactly degenerate eigenstates of $H_{\mathrm{2D}}$
which span a 2D representation $E$.
However, in terms of three--atom eigenstates of $H$,
the states $\psi^{E,\pm}_{\nu_y}(\vec{r})e^{inz}$
occur if the total angular momentum
$n=\mp 1$ modulo 3 because of Eq.~(\ref{eq:psin_rotation}).

The states
$\psi_{\nu_y}^{A_1}(\vec{r})$ and $\psi_{\nu_y}^{E,\pm}(\vec{r})$ obtained for a given $\nu_y$
belong to different representations $A_1$ and $E$. 
Their quasidegeneracy
is lifted by small couplings neglected in the EBK
approach \cite[Sec.~4.5]{bohigas:PhysRep1993}, and the small energy difference
is resolved  in our finite--element
numerical results, as shown on Fig.~\ref{fig:EBK_levels_TrajA}(c).
Because of Eq.~(\ref{eq:psin_rotation}),
the three--atom states $\psi_{\nu_y}^{A_1}(\vec{r})e^{inz}$ occur if 
$n=0$ modulo 3, so that 
none of the three states $\psi_{\nu_y}^{A_1,E\pm}(\vec{r})e^{inz}$ may occur for the same value of $n$.
They do not reduce to an EBK wavefunction corresponding to
a single classical trajectory.
Instead, Eq.~(\ref{eq:psiEBK_TrajA_A1_E}) shows that they
represent coherent superpositions of the three atoms undergoing
motion near the trajectories $A_0$, $A_1$, and $A_2$.

We now turn to quantum states localized near the periodic trajectories of family $C$.
The two  states $\chi^{E,\pm}_{\nu_l}(\vec{r})$, obtained for a given $\nu_l\neq 0$ modulo 3,
are exactly degenerate.
The three--atom states $\chi^{E,\pm}_{\nu_l}(\vec{r})e^{inz}$
occur for
$n=\mp 1$ modulo 3, and opposite values of $n$ lead to atoms rotating along $C$
in opposite directions.
The two states $\chi^{A_1,A_2}_{\nu_l}(\vec{r})$ obtained for a given $\nu_l=0$ modulo 3
belong to different representations and, hence, are  quasidegenerate:
their small energy difference is shown on Fig.~\ref{fig:EBK_levels_TrajC}(c).
The three--atom
states $\chi^{A_1,A_2}_{\nu_l}(\vec{r})e^{inz}$ may occur for the same value of $n=0$
modulo 3.

\subsection{\label{sec:EBK_wavefunctions} EBK quantization: wavefunctions}

\begin{figure*}
  \includegraphics[width=.45\textwidth]
  {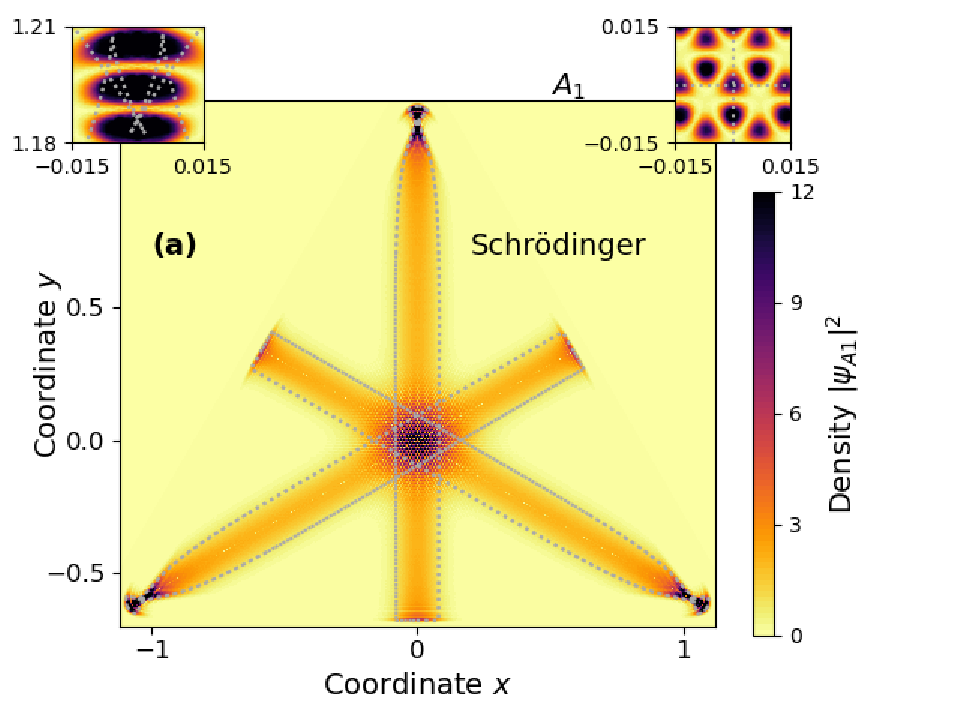}
  \includegraphics[width=.45\textwidth]
  {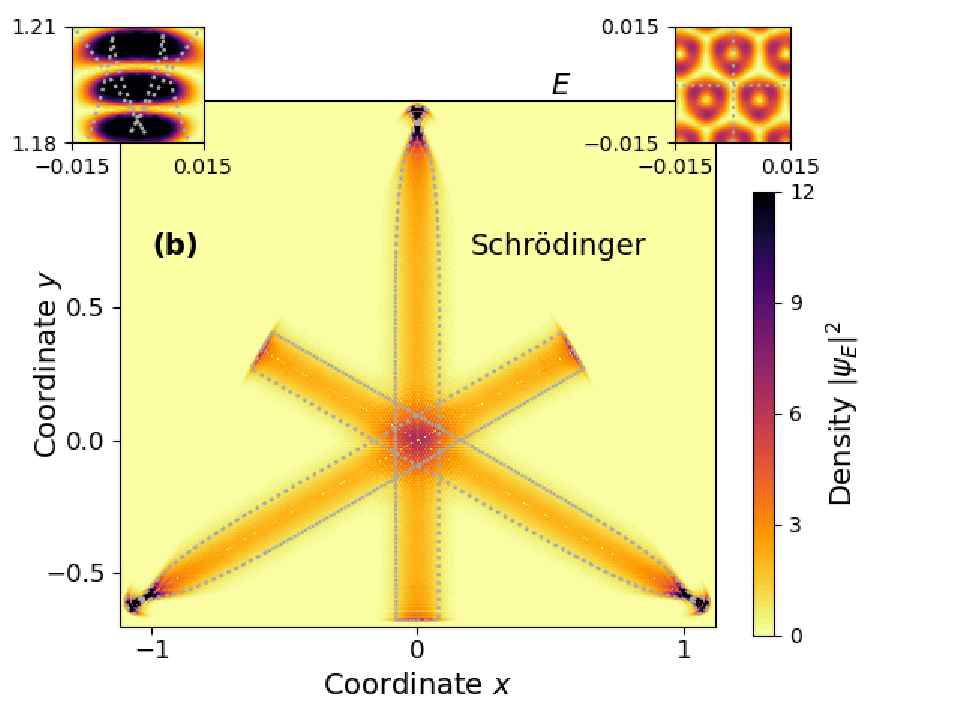}
  \includegraphics[width=.45\textwidth]
  {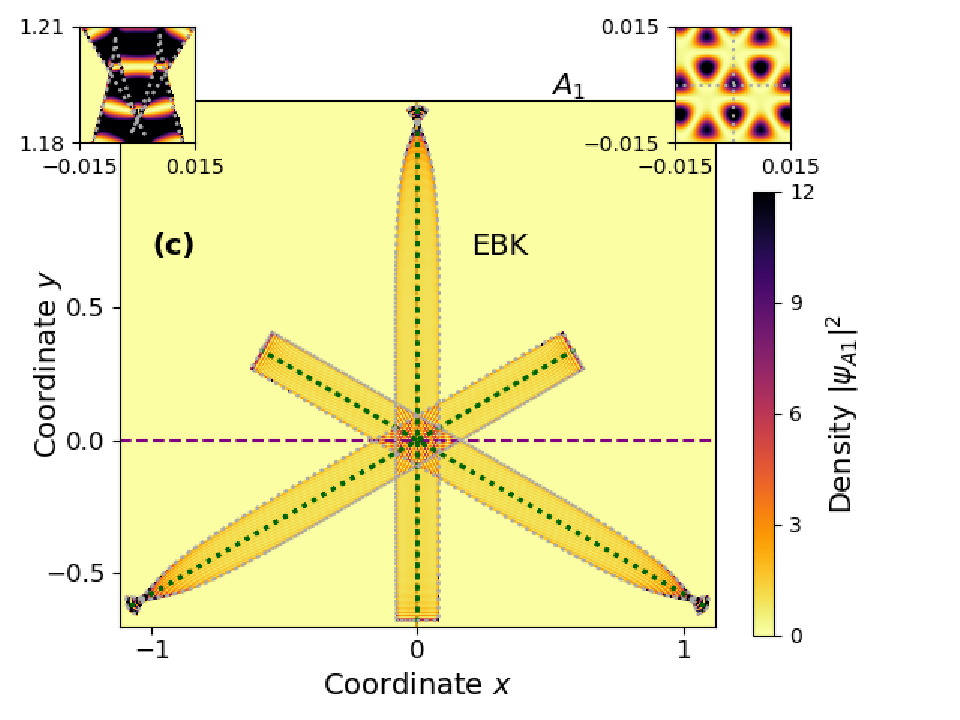}
  \includegraphics[width=.45\textwidth]
  {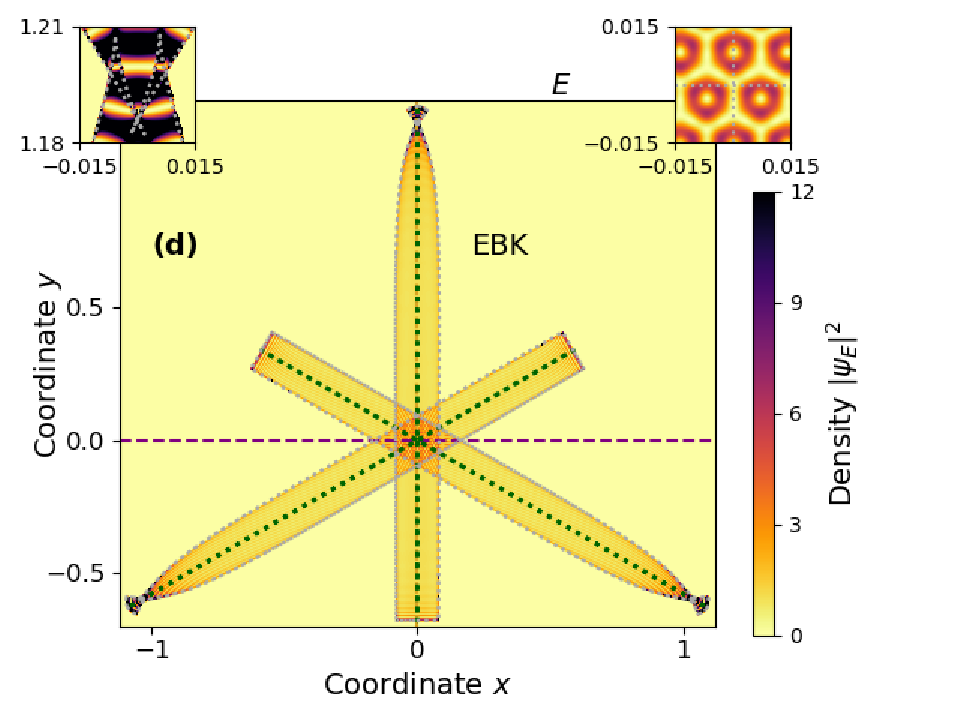}
  \caption{ \label{fig:TrajA_Schrod_EBK}
    \emph{Quantum states localized near the trajectories of family $A$.}
    \textit{{(a,b)}}
    Wavefunction densities $|\psi^{A_1}(\vec{r})|^2$ and $|\psi^{E}(\vec{r})|^2$
    for the two quasidegenerate  eigenstates of $H_\mathrm{2D}$
    localized near the periodic trajectories of family $A$
    whose energies are closest to $C_6/R^6$, obtained through our
    finite--element numerical calculations.
    \textit{{(c,d)}}
    The corresponding squared EBK wavefunctions
    $|\psi_\mathrm{EBK}^{A_1}(\vec{r})|^2$ and $|\psi_\mathrm{EBK}^{E}(\vec{r})|^2$,
    built from the KAM torus satisfying Eq.~(\ref{eq:EBKconds_TrajA})
    with $\nu_x=0$, $\nu_y=300$ (see Fig.~\ref{fig:EBK_levels_TrajA}(a)).
    On all  four panels, the left inset details the region where the
    caustics self--intersect, and the right one shows the region near $(x=0,y=0)$.
  }
\end{figure*}

To further illustrate the applicability of the EBK approach to the
quantum states localized near the stable periodic trajectories of
families $A$ and $C$, we construct primitive EBK wavefunctions
for these states \cite{knudson:JChemPhys1986}.
We focus on a given KAM torus
satisfying the quantization conditions of either Eq.~(\ref{eq:EBKconds_TrajA})
or Eq.~(\ref{eq:EBKconds_TrajC}), depending on whether it lies near a trajectory
of family $A$ or $C$. To obtain the corresponding EBK wavefunctions
$\psi_{\mathrm{EBK}}$ and $\chi_{\mathrm{EBK}}$ of
sections~\ref{sec:energylevels_TrajA} and \ref{sec:energylevels_TrajC} above,
the key extra required step with respect to the approach of
Refs.~\cite{martens:JChemPhys1985,martens:JChemPhys1987} is to describe the
torus in terms of multiple sheets on each of which the classical momentum
is univalued \cite[Sec.~III.A]{percival:AdvChemPhys1977}. These sheets join along the
caustics of the classical trajectory in the $(x,y)$ plane, shown as the dashed
gray lines on Figs.~\ref{fig:EBK_levels_TrajA}(a) and  \ref{fig:EBK_levels_TrajC}(a).
The caustics self--intersect, signalling the occurrence of catastrophes
\cite{delos:JChemPhys1987}, and the torus sheetings must be constructed accordingly.
We find that 12 sheets are required to describe tori near a trajectory of
family $A$ with $\nu_x=0$, and that 6 sheets are required to describe  tori
near a trajectory of family $C$ with $\nu_r=0$. We then obtain
the wavefunctions $\psi_{\mathrm{EBK}}$ and $\chi_{\mathrm{EBK}}$ from the Fourier series
of Eq.~(\ref{eq:condperiodic_Fourierseries}), in terms of linear superpositions
of the contribution of each sheet \cite[III.C]{percival:AdvChemPhys1977}.
Finally, we project $\psi_{\mathrm{EBK}}$ and $\chi_{\mathrm{EBK}}$ onto the irreducible
representations $A_1$, $A_2$, and $E$.

Figure~\ref{fig:TrajA_Schrod_EBK}(c,d) shows
the resulting EBK wavefunctions
for the quasidegenerate quantum states
$\psi^{A_1,E}(\vec{r})$ localized near the trajectories of family $A$
whose energies are closest to $7C_6/R^6$. We compare them to the corresponding wavefunctions
obtained through our finite--element numerical calculations (Fig.~\ref{fig:TrajA_Schrod_EBK}(a,b)).
We show the analogous results for the states $\chi^{A_1,A_2}(\vec{r})$,
localized near the trajectories of family $C$, on figure~\ref{fig:TrajA_Schrod_EBK}.
The agreement between the finite--element and EBK results is excellent,
including in the catastrophe regions where the classical caustics self--intersect,
shown in the upper left insets.

\begin{figure*}
  \includegraphics[width=.45\textwidth]
  {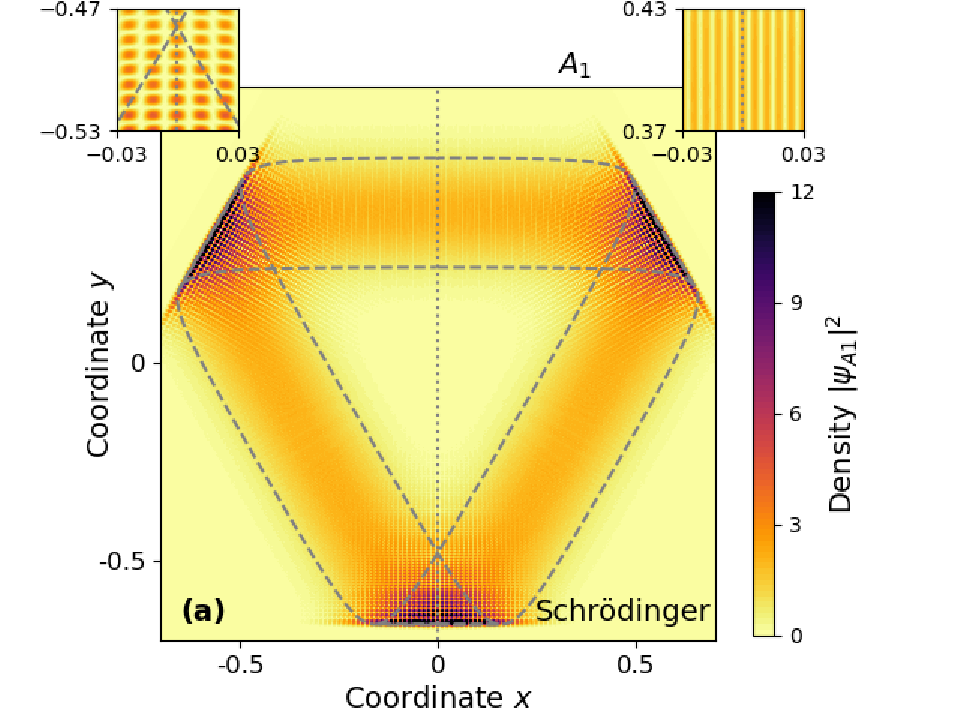}
  \includegraphics[width=.45\textwidth]
  {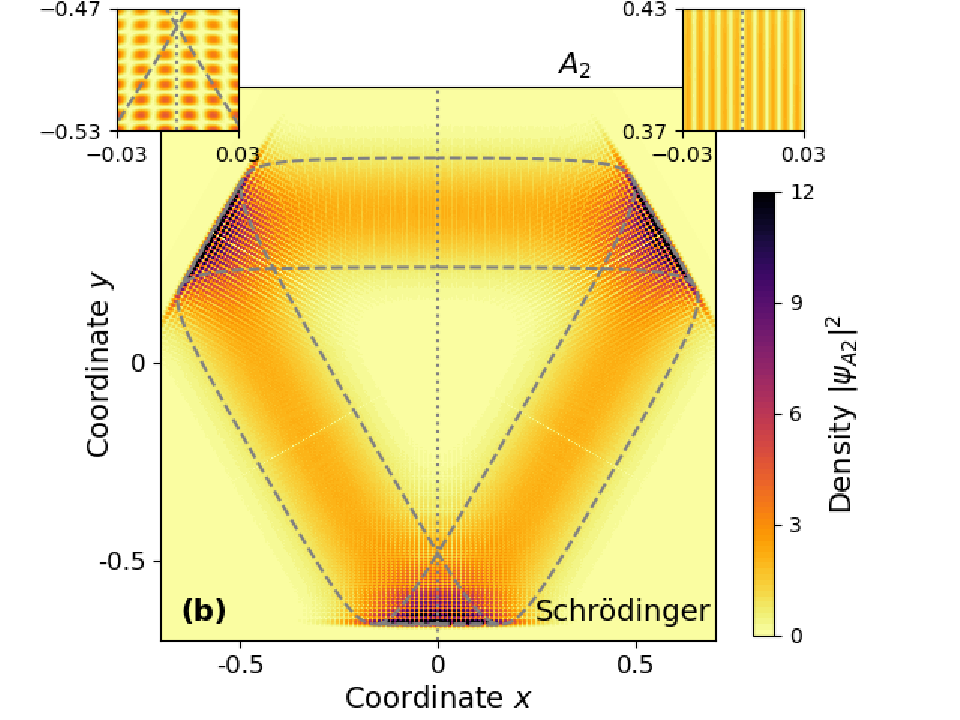}
  \includegraphics[width=.45\textwidth]
  {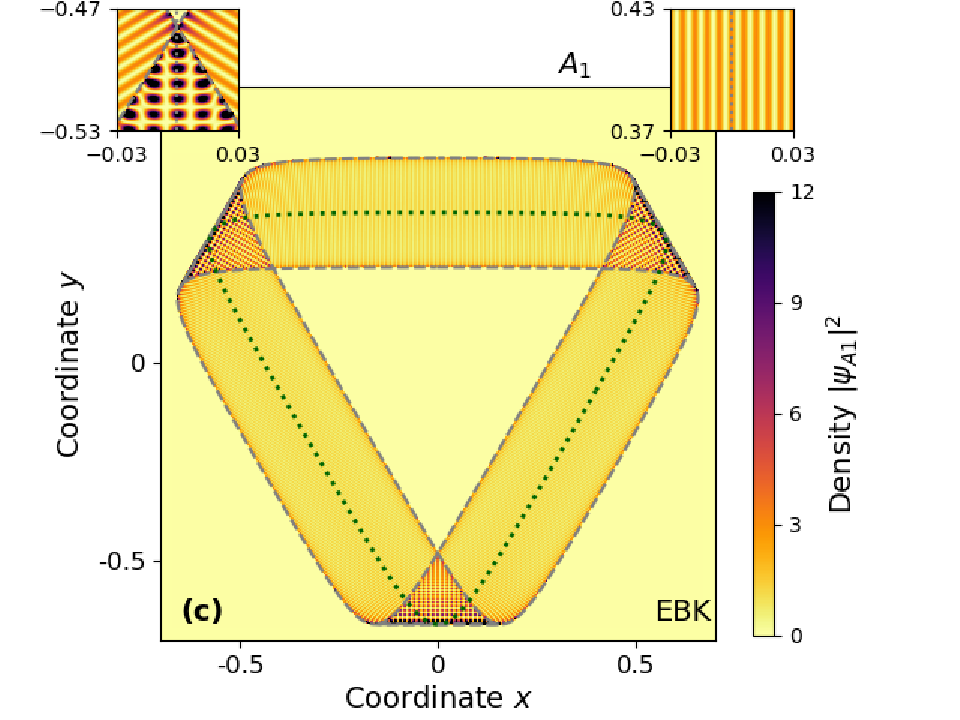}
  \includegraphics[width=.45\textwidth]
  {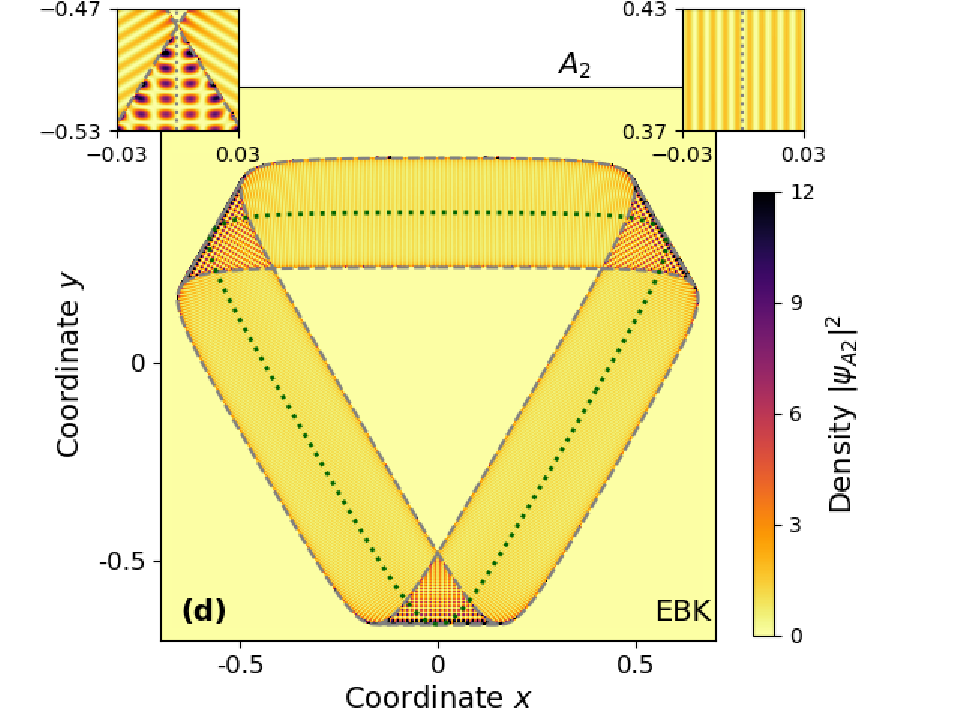}
  \caption{ \label{fig:TrajC_Schrod_EBK}
    \emph{Quantum states localized near the trajectories of family $C$.}
    \textit{{(a,b)}}
    Wavefunction densities $|\chi^{A_1}(\vec{r})|^2$ and $|\chi^{A_2}(\vec{r})|^2$
    for the two quasidegenerate eigenstates of $H_\mathrm{2D}$
    localized near the periodic trajectories of family $C$
    whose energies are closest to $C_6/R^6$, obtained through our
    finite--element numerical calculations.
    \textit{{(c,d)}}
    The corresponding squared EBK wavefunctions
    $|\chi_\mathrm{EBK}^{A_1}(\vec{r})|^2$ and $|\chi_\mathrm{EBK}^{A_2}(\vec{r})|^2$,
    built from the KAM torus satisfying Eq.~(\ref{eq:EBKconds_TrajC})
    with $\nu_r=0$ and $\nu_l=267$ (see Fig.~\ref{fig:EBK_levels_TrajC}(a)).
    On all four panels, the left inset details the region where the caustics
    self--intersect, and the right one shows the region near $(x=0,y=0.4)$.
  }  
\end{figure*}

Primitive EBK wavefunctions do not 
account for the quantum penetration of the wavefunctions through the caustics. 
Instead, they diverge along the caustics as in the WKB approach
\cite[\S 46]{landau3:BH1977}
and vanish outside the classical torus, as illustrated on
Figs.~\ref{fig:TrajA_slices} and \ref{fig:TrajC_slices} in the appendix.
This causes the two limitations of the EBK wavefunctions considered here.
First, interference phenomena involving decaying waves
outside the torus are not captured:
the top left insets of Fig.~\ref{fig:TrajC_Schrod_EBK}
provide an example. Second, the divergence of the wavefunctions leads
to numerical inaccuracies near the caustics which hinder their normalization.
Hence, each of our EBK wavefunctions matches the finite--element wavefunction
up to an overall normalization factor of order 2.
We eliminate it by scaling the EBK wavefunction so that it matches the finite--element
result at one single point chosen far from the caustics.
The quantum penetration through the caustics may be accounted for, and hence
both limitations be overcome, using a uniform approximation to the
wavefunction \cite[Sec.~7.2]{ozoriodealmeida:Cambridge1988}. This goes beyond
the scope of the present work.

\section{ \label{sec:expprospects}
  Experimental prospects and outlook}
The effects considered here may be realized
e.g.\ on the system already considered in Ref.~\cite{papoular:PRA2023}:
${}^{87}\mathrm{Rb}$ atoms in the circular Rydberg
state $50C$, for which $C_6/h=3\,\mathrm{GHz}\,\mathrm{\mu m}^6$.
Then,
the value $\eta=0.01$ is achieved in
a circular trap of radius $R=7\,\mathrm{\mu m}$.
The energy $\epsilon=7C_6/R^6=h\times 200\,\mathrm{kHz}$ is
within experimental reach. For these parameters, the periodic trajectories
of families $A$,
$B$, and $C$ all have periods of the order of $1\,\mathrm{ms}$.
The position of the atoms
may be detected at a given time by turning on a 2D optical lattice
to freeze the dynamics, followed by atomic deexcitation
and site--resolved ground state imaging.
We focus on realizations where the total
three--atom angular momentum
$n$ is well defined.

A key difference between the quantum scar of Ref.~\cite{papoular:PRA2023}
and the localization near stable orbits considered here concerns the
timescale over which quantum particles follow the classical
periodic trajectories.
For the quantum scar, the timescale over which quantum particles
follow the classically unstable periodic trajectory is expected
to depend on its inverse Lyapunov exponent 
\cite[ch.~22]{heller:Princeton2018}.
No such constraint exists
for the dynamics near a classically stable orbit, so that recurrences
of the initial state may be
sought for over the lifetime of the trapped atoms.

Next, we point out a consequence of quantum coherence.
According to Sec.~\ref{sec:role_angularmomentum},
the quantum states
localized near the trajectories of family $A$
are equal--weight superpositions of states localized near the three
periodic trajectories of family $A$ (rather than just one trajectory).
This is the impact of bosonic symmetry.
By contrast,
motion along a single  trajectory $C_+$ or $C_-$
may be observed.

The following point warrants further investigation.
Three atoms launched with angular momentum $n=0$ modulo 3
near the periodic trajectory $C_+$ 
may undergo dynamical tunneling \cite{tomsovic:PhysScr2001}
to the trajectory $C_-$.
The expected oscillation period,
set by
$h/(\epsilon_{\nu_l,A_2}-\epsilon_{\nu_l,A_1})$,
is $\sim 25\,\mathrm{s}$
for the parameters of Fig.~\ref{fig:EBK_levels_TrajC}(c).
This very long timescale is out of reach of current setups, but should become
accessible in
new experiments currently under construction promising atomic lifetimes
$\sim 1\,\mathrm{minute}$ \cite{nguyen:PRX2018,mehaignerie:PRA2023}.
Furthermore, the period may be minimized by varying 
the energy  $\epsilon$ and the parameter $\eta$.
Dynamical tunneling has already been observed for non--interacting,
periodically--driven atoms \cite{hensinger:Nature2001,steck:Science2001}.
The system we consider would provide an example involving
interacting atoms described by a time--independent Hamiltonian.

\section{\label{sec:conclusion}
  Conclusion}

We have revisited the system of three interacting bosonic
particles in a circular trap
that we had first considered in Ref.~\cite{papoular:PRA2023}. We have illustrated
the mixed nature of its classical phase space, and shown
that the statistics of the quantum levels
are well described by a Berry--Robnik distribution. We have analyzed the symmetries 
of the quantum states localized along the classically stable periodic trajectories $A$
and $C$, calculated their energies semiclassically using EBK theory, and constructed
the corresponding EBK wavefunctions. Our semiclassical EBK results,
regarding both the energies
and the wavefunctions, are in excellent agreement with the quantum eigenstates
and energies which we have obtained through finite--element numerical calculations.
Thus, the considered system hosts both a quantum scar, analyzed in Ref.~\cite{papoular:PRA2023},
and classical localization near stable periodic orbits, analyzed in the present work.
These phenomena, all within experimental reach, occur in the same energy
range:
to observe one or the other, one simply adapts the initial conditions
so as to launch the three atoms along a classical periodic orbit which is either unstable or stable.
Hence, the system we propose appears promising in view of a detailed experimental comparison
between quantum scars and classically localized states.

\appendix

\section{Comparison between Schr\"odinger and EBK wavefunctions}

\begin{figure*}
  \includegraphics[width=.35\textwidth]
  {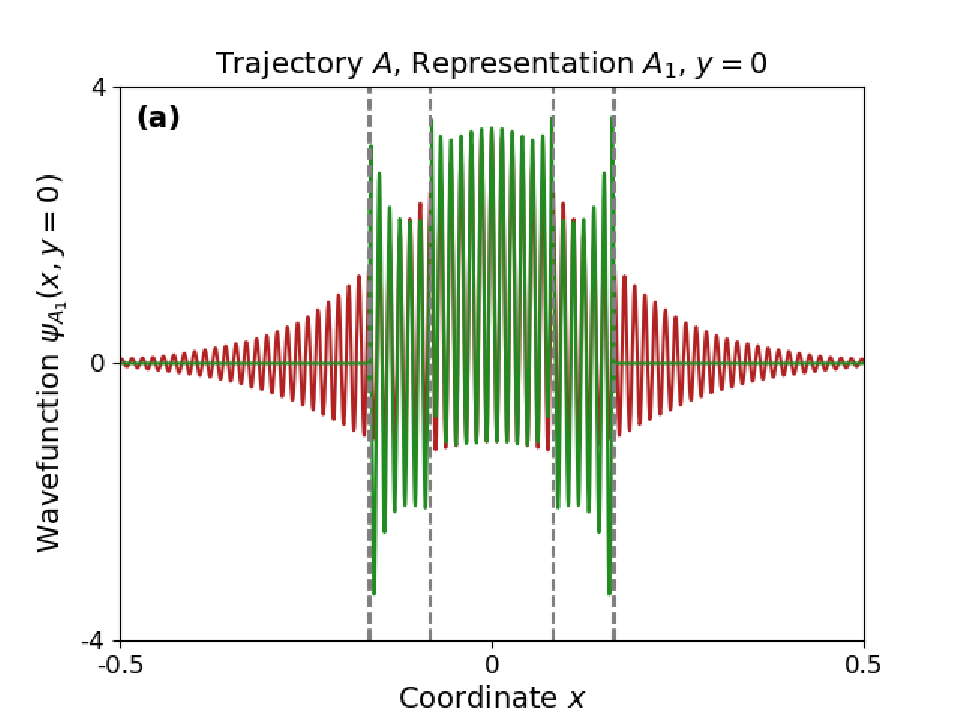}
  \includegraphics[width=.35\textwidth]
  {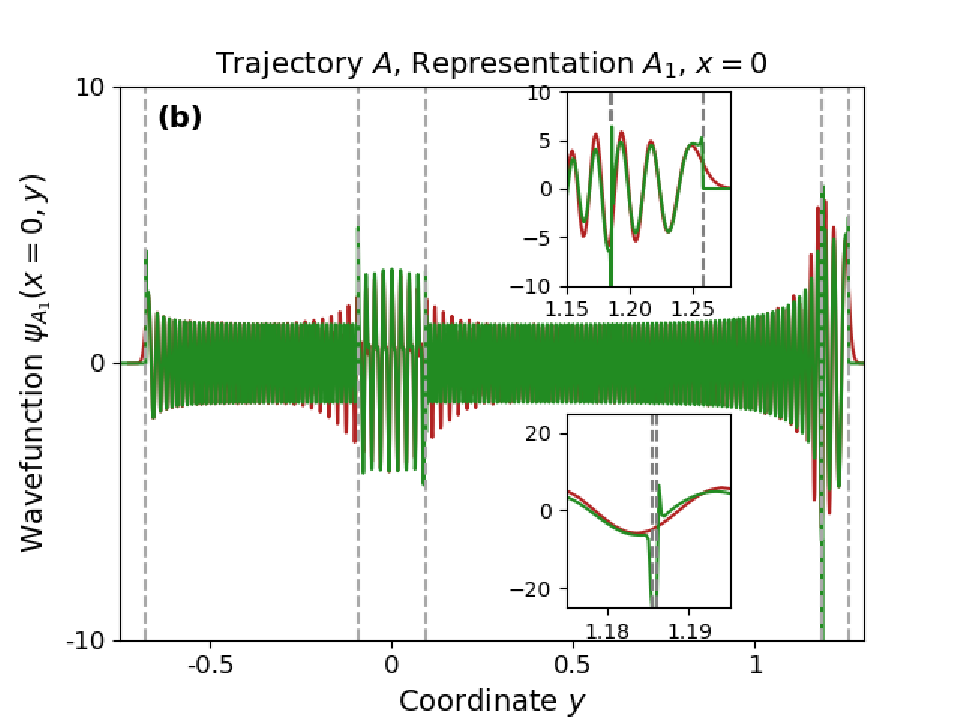}
  \includegraphics[width=.35\textwidth]
  {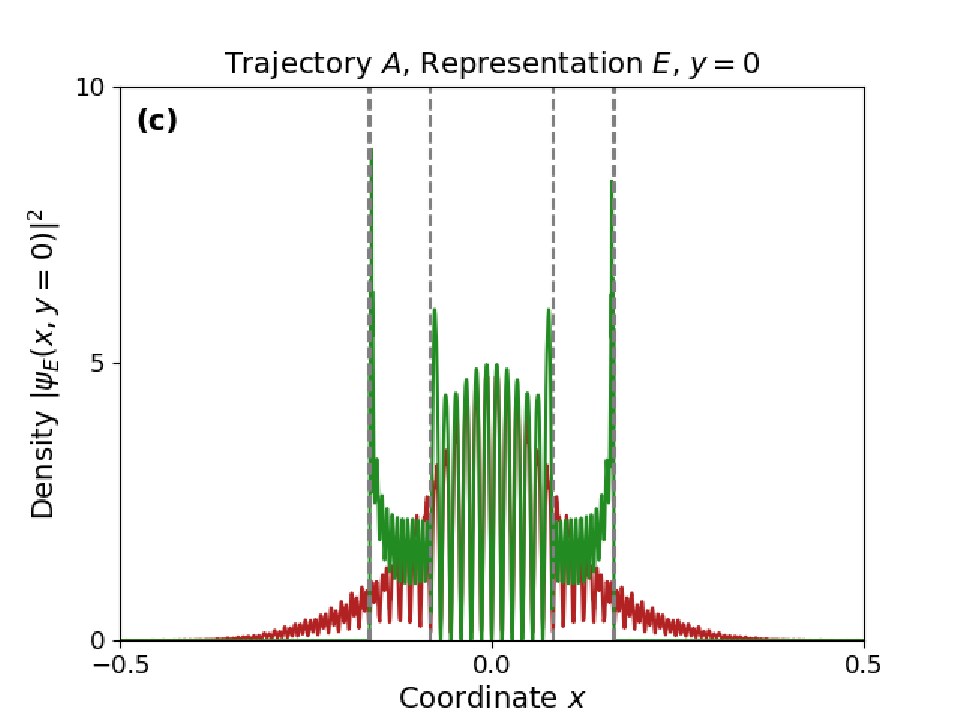}
  \includegraphics[width=.35\textwidth]
  {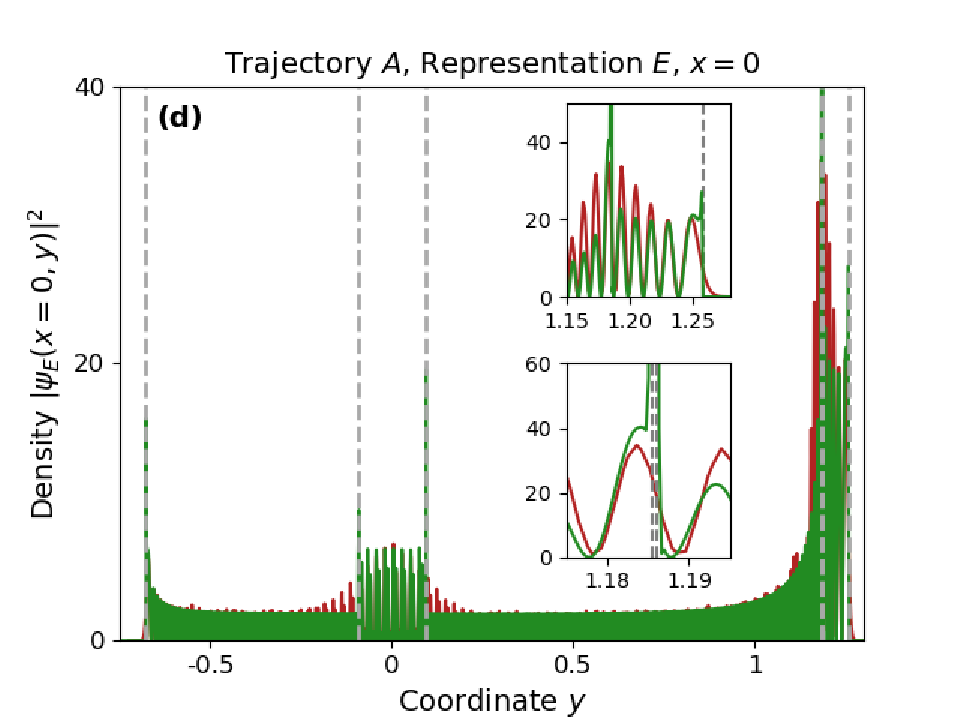}
  \caption{
    \label{fig:TrajA_slices}
    \emph{Quantum states localized near the trajectories of family $A$.} 
    Comparison of the EBK \textit{{(a,b)}}  wavefunction $\psi^{A_1}_{\mathrm{EBK}}$
    and \textit{{(c,d)}} density $|\psi^{E}_{\mathrm{EBK}}|^2$ (green)
    with the corresponding  quantities obtained through finite--element numerics (red)
    shown on Fig.~\ref{fig:EBK_levels_TrajA}, along the horizontal
    \textit{{(a,c)}} and
    vertical \textit{{(b,d)}} axes. 
    The insets illustrate their behaviour near the caustics (vertical dashed gray lines).
    Each EBK wavefunction has been
    scaled to match the finite--element wavefunction at the point $(x=0,y=0.5)$.
  }
  \includegraphics[width=.35\textwidth]
  {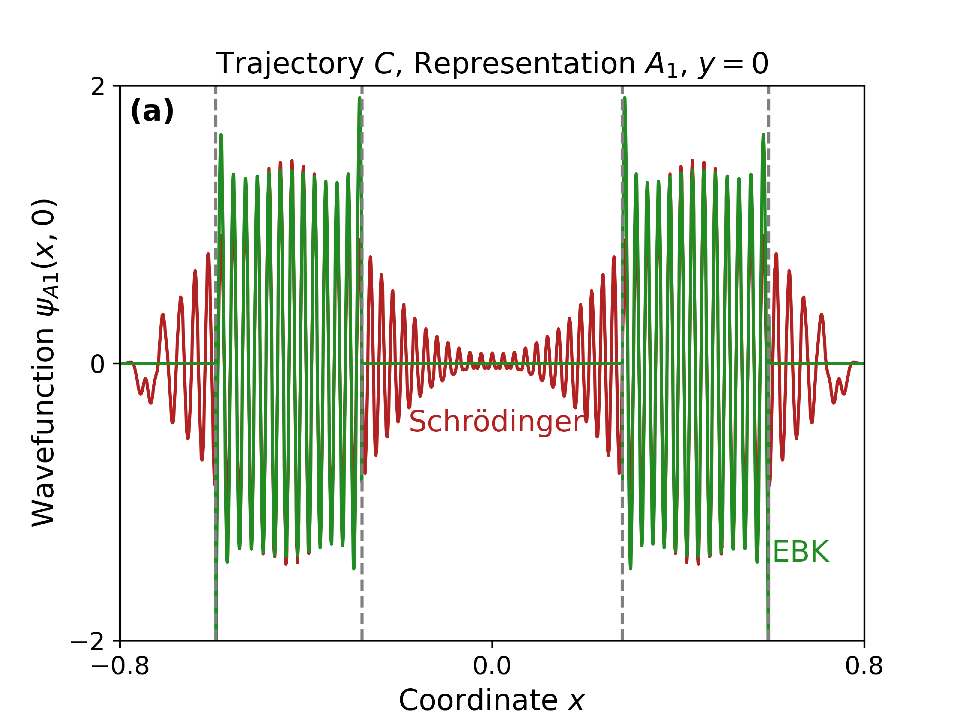}
  \includegraphics[width=.35\textwidth]
  {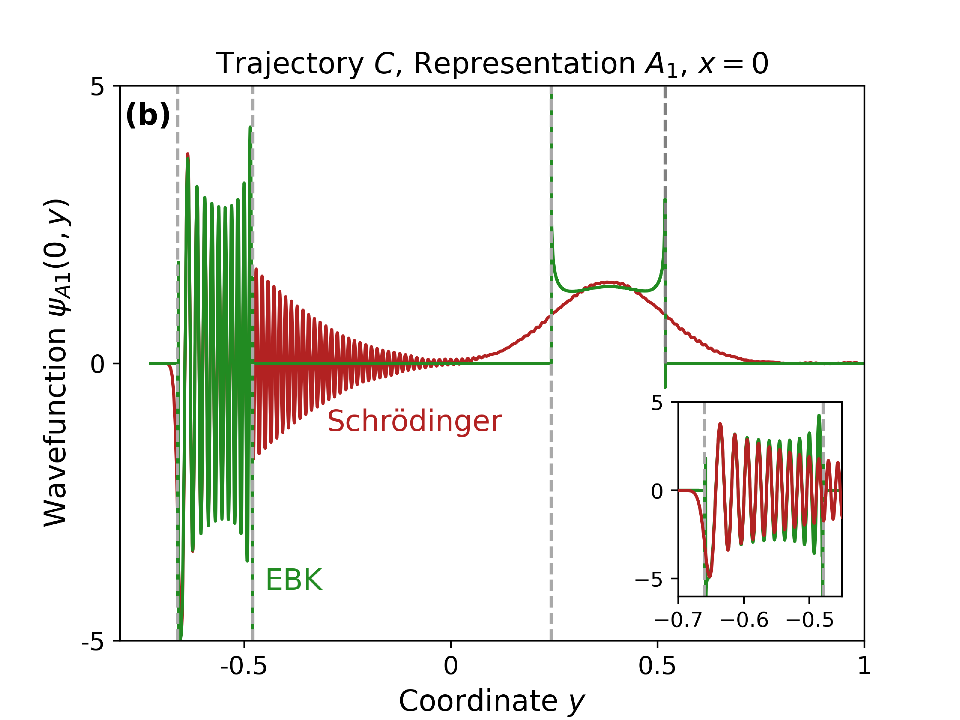}
  \includegraphics[width=.35\textwidth]
  {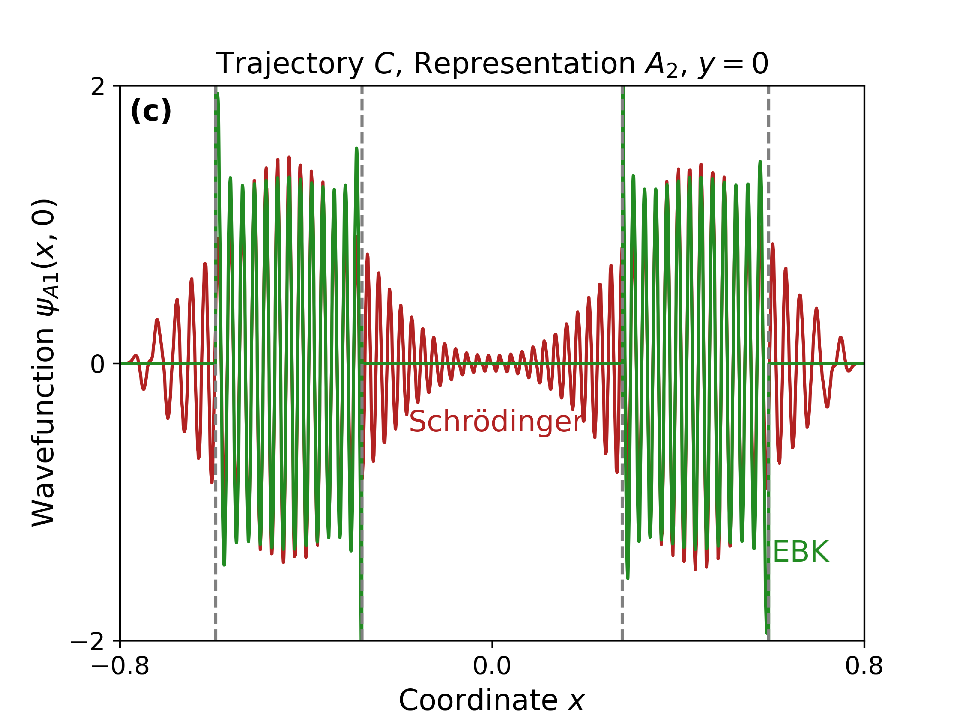}
  \includegraphics[width=.35\textwidth]
  {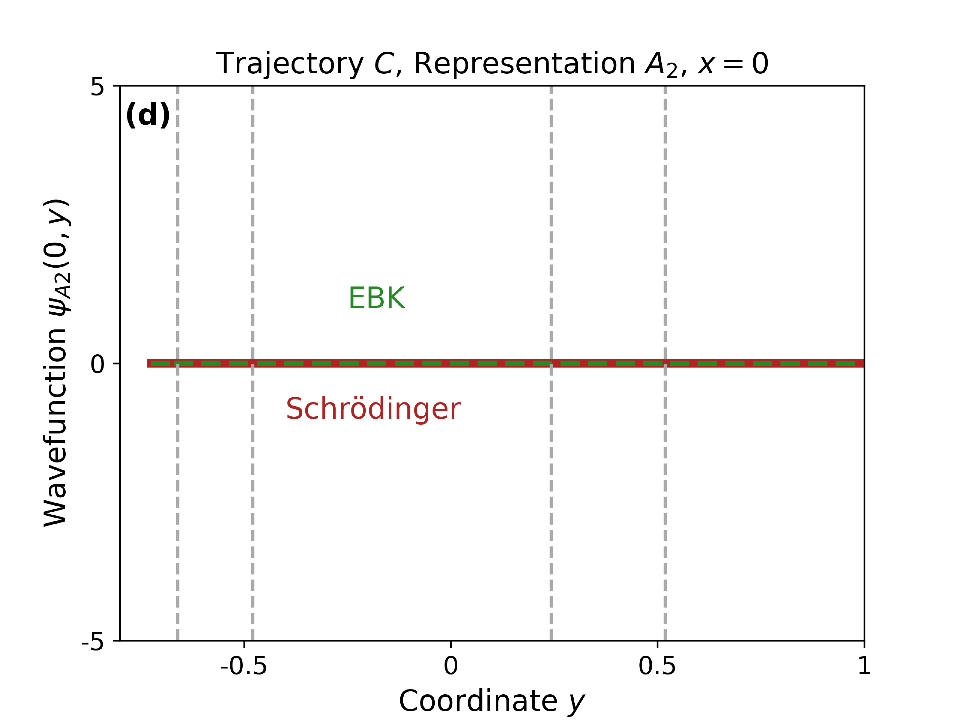}
  \caption{
    \label{fig:TrajC_slices}
    \emph{Quantum states localized near the trajectories of family $C$.}
    Comparison of the EBK wavefunctions (green)
    \textit{(a,b)} $\chi^{A_1}_{\mathrm{EBK}}$
    and \textit{(c,d)} $\chi^{A_2}_{\mathrm{EBK}}$
    and the corresponding wavefunctions obtained through finite--element numerics (red)
    shown on
    Fig.~\ref{fig:EBK_levels_TrajC}, along the horizontal $(a,c)$ and
    vertical $(b,d)$ axes. 
    The insets illustrate their behaviour near the caustics (vertical dashed gray lines).
    Each EBK wavefunction has been
    scaled to match the finite--element wavefunction at the point $(x=0.5,y=0)$.
  }
\end{figure*}

The supplementary figures \ref{fig:TrajA_slices} and \ref{fig:TrajC_slices}
on the next page
compare the behavior of the EBK wavefunctions
to those obtained by solving the Schrödinger equation for the Hamiltonian $H_{\mathrm{2D}}$
through finite--element numerics along the horizontal and vertical axes.
They show excellent agreement between the two approaches,
and highlight the key limitation of the EBK wavefunctions: the quantum penetration
through the caustics is not accounted for, and is replaced by a divergence along the caustics.

\begin{acknowledgments}
  We acknowledge  stimulating discussions with
  M.~Brune and J.M.~Raimond (LKB, Collège de France),
  F.~Dunlop (LPTM, Cergy--Pontoise), and
  R.J.~Papoular (IRAMIS, CEA Saclay).
\end{acknowledgments}

%

\end{document}